\documentclass[11pt]{article}

\usepackage[a4paper,margin=0.82in]{geometry}
\usepackage{amsmath,amssymb}
\usepackage{array}
\usepackage{booktabs}
\usepackage{subfigure}
\usepackage{graphicx}
\usepackage{tabularx}
\usepackage{multirow}
\usepackage{multicol}
\usepackage{pifont}
\usepackage[table,xcdraw]{xcolor}
\usepackage{url}
\usepackage[hidelinks]{hyperref}
\usepackage[title,titletoc]{appendix}

\IfFileExists{microtype.sty}{\usepackage{microtype}}{}

\hypersetup{
    colorlinks=true,
    citecolor=blue,
    filecolor=black,
    linkcolor=red,
    urlcolor=red,
    pdfhighlight =/O
}
\usepackage{subcaption}
\usepackage[most]{tcolorbox}
\usepackage{listings}
\definecolor{instructionframe}{RGB}{77, 101, 138}
\definecolor{instructionback}{RGB}{247, 249, 253}
\definecolor{instructiontitle}{RGB}{230, 236, 246}

\newtcblisting{instructionbox}[1]{%
  enhanced,
  breakable,
  listing only,
  listing engine=listings,
  title={#1},
  colback=instructionback,
  colframe=instructionframe,
  colbacktitle=instructiontitle,
  coltitle=black,
  fonttitle=\bfseries\small,
  boxrule=1.0pt,
  arc=1.5mm,
  left=2.5mm,
  right=2.5mm,
  top=1.0mm,
  bottom=1.0mm,
  before skip=4pt,
  after skip=4pt,
  listing options={
    basicstyle=\ttfamily\small,
    columns=fullflexible,
    breaklines=true,
    breakatwhitespace=true,
    breakindent=0pt,
    breakautoindent=false,
    keepspaces=true,
    showstringspaces=false,
    xleftmargin=0pt,
    aboveskip=5pt,
    belowskip=5pt
  }
}

\newtcolorbox{examplebox}[1]{%
  enhanced,
  breakable,
  title={#1},
  colback=instructionback,
  colbacklower=white,
  colframe=instructionframe,
  colbacktitle=instructiontitle,
  coltitle=black,
  fonttitle=\bfseries\small,
  fontupper=\small,
  boxrule=1.0pt,
  arc=1.5mm,
  left=2.5mm,
  right=2.5mm,
  top=2.0mm,
  bottom=2.0mm,
  before skip=4pt,
  after skip=4pt,
  segmentation style={
    solid,
    draw=instructionframe,
    line width=0.5pt
  },
  before upper={
    \setlength{\parindent}{0pt}
  }
}

\graphicspath{{figures/}}
\newcommand{\QwenThree}{Qwen2.5-VL-3B}
\newcommand{\QwenSeven}{Qwen2.5-VL-7B}
\newcommand{\PaliGemma}{PaliGemma2-3B}
\newcommand{\GemmaFour}{Gemma4-E4B}
\newcommand{\Tsmall}{T50k}
\newcommand{\Tmedium}{T100k}
\newcommand{\Tlarge}{T200k}
\newcommand{\WZthirty}{WZ-30k}
\newcommand{\WZforty}{WZ-40k}
\newcommand{\test}{Test24k}
\newcommand{\pt}{p_{\mathrm{T}}}
\newcommand{\deta}{\Delta\eta}
\newcommand{\dphi}{\Delta\phi}
\newcommand{\JSON}{\textsc{json}}

\newcolumntype{Y}{>{\raggedright\arraybackslash}X}
\newcolumntype{C}[1]{>{\centering\arraybackslash}p{#1}}

\title{
Multitask Jet Analysis with Vision-Language Models: \\A Physics-Informed Four-Panel Representation
}
\author{
Lu Zhang$^{1}$, 
Rachik Soualah$^{1,}$\thanks{Corresponding author: \href{mailto:rachik.soualah@ku.ac.ae}{rachik.soualah@ku.ac.ae}}, 
Abbes Amira$^{2}$
\\[0.35em]
\small 
$^{1}$Department of Physics, Khalifa University of Science and Technology, \\
\small P.O. Box 127788, Abu Dhabi, United Arab Emirates\\
\small 
$^{2}$Faculty of Science and Engineering, University of Wolverhampton, \\
\small WV1 1LY, Wolverhampton, United Kingdom
}
\date{\today}

\begin{document}

\maketitle

\begin{abstract}

The enormous recorded data at high-energy physics (HEP) colliders makes the accurate identification of physics objects a bottleneck in disentangling event topologies, where machine learning has become a standard tool for jet tagging. In this work, we examine whether Vision-Language Models (VLMs) can provide a common interface for structured jet analysis from a single physics-informed image. Each JetClass jet becomes a $224\times224$ RGB image with four panels encoding $p_T$ flow of all constituents; charged-hadron, neutral-hadron, and electromagnetic composition; $p_T$-weighted impact-parameter significances with displaced-track multiplicity; and signed-track $p_T$ densities with local $p_T^{\kappa}$-weighted jet-charge asymmetry. Four open VLMs are adapted with low-rank adaptation (LoRA) for ten-class classification across QCD, Higgs, W/Z, and top jets, six-field attribute prediction, and cross-panel consistency with replaced-panel localization, evaluated on 24000 balanced JetClass test jets. Ablations identify the impact-parameter lifetime signature of heavy-flavor tagging as dominant, with jet charge separating the nearly mass-degenerate hadronic W and Z. Zero-shot recall stays near chance ($3.33\%$--$10.57\%$); once adapted, Task~1 Macro recall grows monotonically with training size, and Gemma4-E4B reaches $75.05\%$ ($74.87\%$ $F_1$), $93.31\%$ Task~2 field-mean, and $99.88\%/99.75\%$ binary/localization recall. Reallocating budget toward W/Z jets raises \texttt{Zqq} recall by up to $9.83$ points at near-constant Macro recall. Transfer to broader-topology JetClass-II and real-data Aspen Open Jets (probing the simulation-to-data gap) is retained: with 3000 target jets, Task~1 recall exceeds $70\%$ on JetClass-II and every Task~2/3 metric exceeds $92\%$. Thus rasterizing energy flow, species, displacement, and jet charge, with parameter-efficient adaptation, supports jet analysis through one instruction-conditioned interface for future collider data.
\end{abstract}

\section{Introduction}\label{sec:introduction} 
The CERN-Large Hadron Collider (LHC) delivers proton-proton collision datasets of unprecedented volume and complexity, yielding \(\mathcal{O}(10)\,\mathrm{PB}\) per year after online selection, with the expected High-Luminosity upgrade set to increase the integrated dataset by roughly an order of magnitude at an average pileup of \(\sim\!200\) interactions per bunch crossing~\cite{atlas_upgrade_hllhc,atlas_run2_electroweak_qcd_flavour}. Extracting physics from such data hinges on the accurate identification of reconstructed objects in particular jets, whose internal structure encodes the initiating parton as rare signals are isolated from an overwhelmingly larger Standard Model background across event topologies of high final-state multiplicity~\cite{atlas_toward_hllhc_electronics,qu2022particle}. The scale and dimensionality of this problem have made machine learning a central tool for collider analysis, from low-level reconstruction to jet tagging, since the discriminating information in a jet's constituents is difficult to capture with hand-designed observables alone ~\cite{atlas_toward_hllhc_electronics,qu2022particle}.

Jet tagging maps a variable-length collection of reconstructed particles to the physical process associated with a jet. Jets originate in the fragmentation and hadronization of high-energy quarks and gluons and are reconstructed as collimated sprays of hadrons whose internal structure retains information about the initiating parton. At sufficiently high transverse momentum, the hadronic decays of \(W\), \(Z\), Higgs bosons, and top quarks become collimated into single large-radius jets, so that the parent mass, prong multiplicity, and heavy-flavor content are imprinted on the internal constituent distribution rather than resolved as separate jets~\cite{Larkoski:2017jix,Marzani:2019hga}. Separating these signatures from the overwhelmingly more abundant QCD multijet background underlies a broad range of Standard Model measurements and searches for new phenomena, and defines the physics context of the classification task considered here in this presented work.

The Particle Transformer achieves strong performance by operating directly on constituent sets and encoding particle-pair relations~\cite{qu2022particle}. Such specialized architectures are natural choices when the objective is a single discriminative output. A complementary question is whether an open, pretrained Vision--Language Model (VLM) can provide a common interface for multiple jet-analysis tasks, with both the task and the required output specified through language.

Transformer-based visual encoders and large-scale image--text pretraining provide important foundations for modern VLMs~\cite{dosovitskiy2020image, radford2021learning}. Instruction-following VLMs connect visual encoders to language models and can be adapted through multimodal instruction tuning~\cite{liu2023visual, dai2023instructblip}. Their motivation in jet physics is not the expectation that visual pretraining on conventional images already contains collider knowledge. Instead, a task-conditioned generative interface can support classification, structured attribute prediction, and consistency assessment without introducing a separate output head for each task. This setting presents two challenges. First, the physically distinct constituent information must remain accessible despite the intrinsic sparsity of jet images: the angular size of a jet is set by its opening angle, \(\Delta R \sim 2m/p_T\) for a boosted two-body decay, so the discriminating structure occupies a small fraction of the \((\deta,\dphi)\) plane and, at the coarse patch resolution of a VLM encoder, many patches carry no constituent activity while the flavor/ charge-sensitive signals reside in weak, sparsely populated regions. Second, evaluation must separate knowledge acquired through domain adaptation from compliance with a prompt or output schema.

This work introduces a deterministic four-panel representation of the constituents of jets obtained from JetClass which is a large public benchmark of simulated LHC jets built for training machine-learning taggers~\cite{qu2022particle,qu2022jetclass}. All panels share the same centered and canonicalized \((\deta,\dphi)\) coordinate system, while their RGB channels encode complementary information about transverse-momentum composition, particle species, track displacement, and charge structure. The resulting input is a single \(224\times224\) RGB image together with a task-specific instruction. Geometric canonicalization translates each jet to its \(p_T\)-weighted centroid, conditionally rotates it onto its principal (substructure) axis, and applies deterministic reflections to fix the residual discrete symmetries, suppressing nuisance dependence on jet position and orientation while preserving the internal substructure. Finite-width Gaussian deposition and logarithmic intensity compression then counter the sparsity, correlating neighboring pixels and raising the visibility of soft radiation without distorting the relative energy flow within each panel. No auxiliary scalar values are supplied as text or separate numerical input. The physically separated panels make the model input directly inspectable while preserving a common spatial correspondence across all twelve encoded fields. The encoded quantities follow the observables that drive established tagging strategies. The transverse-momentum flow carries the energy-weighted angular structure from which the jet mass and \(N\)-prong substructure are constructed, separating the one-prong QCD background from the two-prong \(W/Z\) and \(H\to q\bar q\) topologies and the three-prong hadronic top~\cite{Thaler:2010tv}. The charged, neutral, and electromagnetic composition reflects the color charge and fragmentation of the initiating parton, for which the gluon-to-quark constituent-multiplicity ratio approaches the Casimir ratio \(C_A/C_F = 9/4\) in the soft  limit~\cite{Gallicchio:2011xq}. Track displacement encodes the lifetime signature of weakly decaying \(b\) and \(c\) hadrons that underlies impact-parameter and secondary-vertex flavor tagging, providing the primary handle on the flavor content distinguishing \(H\to b\bar b\), \(H\to c\bar c\), and the \(b\)-tagged top categories. The charge structure encodes the \(p_T^{\kappa}\)-weighted jet charge, an estimator of the electric charge of the fragmenting system~\cite{Krohn:2012fv}, which supplies the residual discriminant between the nearly mass-degenerate, morphologically similar hadronic \(W^{\pm}\) and \(Z\). Rendering these quantities as spatially aligned image fields exposes, within a single input, the physics that dedicated substructure, flavor, and charge observables otherwise quantify separately.

The main contributions within this work are:
\begin{itemize}
    \item \textbf{Physics-informed visual encoding.}
    A deterministic four-panel jet representation encodes transverse-momentum composition, particle species, track displacement, and charge structure in twelve spatially aligned physical fields distributed across four RGB panels. The resulting \(224\times224\) image is human-inspectable and is supplied without auxiliary jet variables or truth metadata.
    \item \textbf{A single-adapter multitask interface.}
    For each checkpoint, a single LoRA adapter~\cite{hu2022lora} enables the corresponding VLM to perform three structurally different tasks through a shared image--text interface: ten-class truth-category classification, six-field structured attribute prediction, and cross-panel consistency assessment with replaced-panel localization. The three tasks produce nine typed fields through compact \JSON{} schemas without task-specific prediction heads.
    \item \textbf{A controlled evaluation of adaptation and data scale.} 
    A common experimental framework systematically examines the effects of panel content, domain adaptation, multitask supervision, training-set size, and class allocation. Four open VLM configurations are evaluated using two zero-shot conditions, panel and multitask ablations, balanced subsets of 50,000--200,000 source jets, and fixed-budget \(W/Z\)-enriched subsets. With 200,000 source jets (only \(0.2\%\) of the official 100-million-jet JetClass training partition~\cite{qu2022jetclass}), the best configuration reaches \(75.05\%\) Task~1 Macro recall, \(93.31\%\) Task~2 field-mean Macro recall, \(99.88\%\) Task~3 binary Macro recall, and \(99.75\%\) localization Macro recall on inconsistent examples.
    \item \textbf{Error-guided fixed-budget data strategy.} 
    Class-wise evaluation identifies \texttt{Zqq} as a persistent performance bottleneck. Increasing the \texttt{Zqq} and \texttt{Wqq} training quotas within a fixed 200,000-jet budget improves \texttt{Zqq} recall by up to \(9.83\) percentage points, demonstrating that targeted class reallocation can improve a difficult category without increasing the total training-set size.
    \item \textbf{A unified end-to-end inference workflow.} 
    A common inference workflow is implemented for all four adapted VLM configurations. Given a four-panel test image and a compatible LoRA adapter, it executes the three task-specific instructions, validates the generated \JSON{} objects, and records the predictions together with model and adapter provenance.
    \item \textbf{Cross-dataset transfer with limited adaptation.} 
    Generalization to JetClass-II~\cite{li2024accelerating,jetclass2} and Aspen Open Jets~\cite{amram2025aspen} is evaluated using random baselines, frozen transfer of the JetClass-trained adapters, and lightweight target-domain adaptation with only 3,000 training jets and 600 validation jets. The results show that structured attribute prediction and panel-consistency assessment transfer effectively through the shared four-panel representation, whereas truth-category classification benefits from limited adaptation to the new label space.
\end{itemize}

The outline of the paper is organized as follows. Section~\ref{sec:related} reviews jet representations, physics foundation models, and instruction-tuned VLMs from previous related works by other groups (described below). Sections~\ref{sec:data_representation} and \ref{sec:tasks} define the four-panel input and the three instruction tasks. Section~\ref{sec:setup} describes model adaptation and evaluation. Section~\ref{sec:results} first presents the panel and multitask ablations, then examines zero-shot performance and balanced-data scaling, and finally studies fixed-budget \(W/Z\) enrichment and cross-dataset transfer learning. Section~\ref{sec:conclusion} gives the conclusion of the paper.
\section{Related Work}\label{sec:related}
Two representation families are particularly relevant to the present study: constituent-based and image-based approaches~\cite{kheddar2024image}. Constituent-based approaches retain individual reconstructed particles and process them as sets, point clouds, graphs, or sequences. 
Energy Flow Networks (EFNs) and Particle Flow Networks (PFNs) learn directly from unordered particle sets through a per-constituent latent map followed by a symmetric aggregation~\cite{komiske2019energy}; in the EFN the constituents enter only through energy-weighted angular functions, so the observable is infrared and collinear (IRC) safe by construction and calculable in perturbative QCD. ParticleNet treats a jet as a permutation-invariant particle cloud and applies dynamic graph convolutions over learned nearest neighbors in the \((\eta,\phi)\) plane~\cite{qu2020jet}, while LorentzNet enforces equivariance under the proper orthochronous Lorentz group through message passing built from Minkowski inner products~\cite{gong2022efficient}. Particle Transformer injects pairwise interaction features into its attention, constructed from Lorentz-invariant quantities of each constituent pair their angular separation, relative transverse momentum, and invariant mass~\cite{qu2022particle}. These approaches preserve the variable-length constituent information and encode the relevant space-time and gauge symmetries, but they depend on architectures tailored to sets, graphs, or sequences. Their inputs remain physically meaningful, yet do not form a fixed visual object that can be passed unchanged to a pretrained visual encoder.

Image-based approaches rasterize calorimeter or particle-flow activity on a fixed angular grid~\cite{kagan2022image}. Jet images mapped calorimeter deposits in the \((\eta,\phi)\) plane to pixel intensities, the pixelization following the calorimeter segmentation in pseudorapidity and azimuth~\cite{cogan2015jet, de2016jet}. Multi-channel ``color'' images then split the input into charged transverse momentum, neutral transverse momentum, and charged-particle multiplicity~\cite{komiske2017deep}; because gluon jets radiate more than quark jets by the color Casimir ratio \(C_A/C_F = 9/4\), the multiplicity channel is sensitive to the color charge of the initiating parton, and this decomposition was found to improve quark/gluon discrimination over a single-channel image. The resulting fixed-size representation is human-inspectable and readily compatible with convolutional and transformer-based visual encoders. However, the regular grid images can be highly sparse: fine binning produces many empty or weakly populated pixels, whereas coarse binning merges nearby deposits. Pixelization can therefore discard information that remains explicit in a constituent list~\cite{kagan2022image}. 
For patch-based encoders the sparsity is compounded: a jet occupies few patches, and many carry no constituent activity. Earlier jet-image analyses have addressed this problem and reduced it by Gaussian-smoothing the pixelized deposits to correlate neighboring cells~\cite{dillon2023normalized, buss2023anomalous,oleksiyuk2024cluster}. The present representation applies the same principle during rasterization (the binning of constituents onto a regular \((\deta,\dphi)\) grid, with finite-width Gaussian deposition rather than single-pixel assignment). Furthermore, each constituent contribution is deposited with a finite-width Gaussian kernel rather than assigned to a single pixel; the resulting correlation of adjacent pixels is an image-space analogue of the approximate collinear stability of a well-behaved jet observable. Logarithmic intensity compression, likewise with precedent in jet-image preprocessing~\cite{komiske2017deep}, suppresses the hardest deposits and raises the relative visibility of soft radiation, after which per-panel normalization maps the result onto the available intensity range. These steps do not recover the constituent-level information lost to pixelization, but they yield a denser, more accessible representation at the fixed resolution imposed by general-purpose VLMs. In addition, the four-panel layout distributes twelve physics-defined image fields across four spatially aligned RGB panels, allowing transverse-momentum composition, particle species, track displacement, and charge structure to remain visually separated while preserving a standard single-image input.

Recent collider-physics research has moved beyond task-specific taggers toward reusable pretrained representations. Early self-supervised work such as JetCLR learned jet representations by contrastive learning with augmentations drawn from physical jet symmetries-azimuthal rotations, \((\eta,\phi)\) translations, and soft or collinear splittings~\cite{dillon2022symmetries}. Masked particle modeling learns permutation-invariant set representations by reconstructing masked constituents and supports transfer to several downstream jet tasks~\cite{golling2024masked}. OmniJet-\(\alpha\) uses a tokenized particle representation and demonstrates transfer from jet generation to jet tagging~\cite{birk2024omnijet}. Subsequent work has shown that continuous reconstruction objectives can avoid discrete particle tokenization~\cite{leigh2025tokenization}. Other studies have examined jet foundation models across multiple collider tasks and large-scale self-supervised pretraining to improve label efficiency~\cite{mikuni2025solving, zhao2024large}. Aspen Open Jets provides large samples derived from CMS open collision data for such studies~\cite{amram2025aspen}, while HEP-JEPA learns latent jet representations through joint-embedding prediction~\cite{bardhan2025hep}. Despite their different objectives, these approaches share a physics-native strategy: a dedicated encoder is pretrained directly on constituent-level data, with the architecture and learning objective designed around the structure of particle events. These constituent-level approaches retain direct access to fine-grained particle information but generally rely on domain-specific encoders and pretraining objectives.

Visual instruction tuning provides a complementary approach, namely reusing a general multimodal model instead of pretraining a new physics-specific backbone model. LLaVA demonstrated end-to-end visual instruction tuning by connecting a pretrained visual encoder to a language model~\cite{liu2023visual}, and InstructBLIP showed that a natural-language instruction can condition the extraction of task-relevant visual features~\cite{dai2023instructblip}. LoRA further enables parameter-efficient transfer while keeping the pretrained model weights frozen~\cite{hu2022lora}. However, general-purpose VLMs are not designed specifically for sparse particle-flow maps. Performance on natural images, documents, or conventional vision benchmarks does not guarantee recognition of physics-defined RGB channels. Particle-flow maps, unlike natural scenes, carry no photometric texture or object contours; their channels encode physical densities-energy flow, particle species, track displacement, and charge-rather than visible color, and the priors from large-scale image pretraining therefore do not transfer directly. Furthermore, a recent study adapted an LLaMA-based VLM to classify neutrino interactions from paired orthogonal detector views and compared it with CNN and Vision Transformer baselines~\cite{sagar2026adapting}, which provides an early demonstration that VLMs can be adapted to sparse scientific event displays. However, applications of instruction-following VLMs to particle-physics imagery remain relatively limited. 
Accordingly, public VLM checkpoints are adapted with LoRA to interpret a single physics-informed four-panel image. The aim is not to supplant constituent-based foundation models, which operate on the full set of particle four-momenta and are purpose-built for jet physics, but to test whether a general multimodal model can learn the semantics of human-inspectable, physics-defined image fields from at most 200,000 source jets only \(0.2\%\) of the official JetClass training partition~\cite{qu2022jetclass} and serve multiple instruction-defined tasks within one model. Such a capability would provide a complementary interface for jet analysis without requiring a separate architecture or prediction head for each task.

Multitask learning is attractive because a shared model can reuse representations across related objectives, reduce duplicated model components, and provide a unified inference interface. However, the gains are not guaranteed: task imbalance and incompatible optimization dynamics can reduce or reverse them~\cite{caruana1997multitask, chen2018gradnorm}. In jet physics, OmniLearn showed that one constituent-level representation can serve jet tagging, event generation, likelihood-ratio estimation, and anomaly detection, and transfer across samples with different detector or collision conditions~\cite{mikuni2025omnilearn}, a physics-native precedent for cross-task reuse. The present study takes a complementary, instruction-based route: the task is specified in language, and a common autoregressive decoder emits the structured output without task-specific heads. Each LoRA-adapted VLM reads the same four-panel jet image under three
instructions and returns compact \JSON{} objects for ten-class truth-category classification across QCD, \(W/Z\), Higgs, and top jets; substructure-attribute prediction; and cross-panel consistency assessment with replaced-panel localization. This motivates the central question: whether a general-purpose VLM can learn a physics-informed four-panel representation and support heterogeneous structured tasks through a shared instruction interface.
\section{Dataset and Four-Panel Representation}\label{sec:data_representation}
\subsection{JetClass samples and experimental subsets}
This study uses jets from JetClass, a simulated benchmark released with the Particle Transformer study~\cite{qu2022particle, qu2022jetclass}. The official release contains ten equally populated categories and is partitioned into 100 million training jets, 5 million validation jets, and 20 million test jets. Event generation uses \textsc{MadGraph5\_aMC@NLO}, followed by parton showering and hadronization with \textsc{Pythia} and detector simulation with \textsc{Delphes}. Jets are clustered with the anti-\(k_{\mathrm T}\) algorithm using \(R=0.8\), and satisfy \(500 < \pt < 1000~\mathrm{GeV}\) and \(|\eta|<2\). Table~\ref{tab:classes} lists the ten Monte Carlo truth categories.

\begin{table}[!htpb]\centering\footnotesize
\renewcommand{\arraystretch}{1.0}
\caption{Labels used by the JetClass data~\cite{qu2022particle, qu2022jetclass}. Each label denotes a Monte Carlo truth category. Charge-conjugate modes are implicit, and \(\ell\in\{e,\mu\}\).}
\label{tab:classes}
\begin{tabular}{
  >{\centering\arraybackslash}p{1.5cm}
  >{\centering\arraybackslash}p{6cm}
}
\hline \hline
\textbf{Label}   & \textbf{Monte Carlo Truth Category} \\ 
\hline
\texttt{QCD}     & light-quark or gluon jet (\(q/g\) background) \\
\texttt{Hbb}     & \(H\rightarrow b\bar b\) \\
\texttt{Hcc}     & \(H\rightarrow c\bar c\) \\
\texttt{Hgg}     & \(H\rightarrow gg\) \\
\texttt{H4q}     & \(H\rightarrow 4q\) \\
\texttt{Hqql}    & \(H\rightarrow \ell\nu qq'\) \\
\texttt{Zqq}     & \(Z\rightarrow q\bar q\) \\
\texttt{Wqq}     & \(W\rightarrow qq'\) \\
\texttt{Tbqq}    & \(t\rightarrow bqq'\) \\
\texttt{Tbl}     & \(t\rightarrow b\ell\nu\) \\
\hline \hline
\end{tabular}
\end{table}

All subsets respect the official JetClass partition boundaries. Training jets are drawn exclusively from the official training partition. A fixed, class-balanced subset of 5,000 validation jets is used to monitor the validation loss during adaptation. Final evaluation uses a fixed, class-balanced sample of 24,000 jets drawn exclusively from the official test partition, denoted \test{}.

Unless otherwise stated, dataset sizes refer to distinct source jets before multitask expansion. Each source jet yields one task-specific instance for each of the three tasks, so \(N\) source jets correspond to \(3N\) task instances. The class-balanced subsets \Tsmall{}, \Tmedium{}, and \Tlarge{} contain 5,000, 10,000, and 20,000 training jets per category, respectively. Their total sizes are therefore 50,000, 100,000, and 200,000 source jets, corresponding to 150,000, 300,000, and 600,000 task instances.

To isolate the effect of class allocation from that of total training size, two fixed-budget \(W/Z\)-enriched alternatives to \Tlarge{} are constructed. The \WZthirty{} and \WZforty{} configurations assign 30,000 and 40,000 jets, respectively, to each of the \texttt{Zqq} and \texttt{Wqq} categories; the remaining budget is divided equally among the other eight categories. These configurations reallocate, rather than augment, the 200,000-jet \Tlarge{} budget. The exact class quotas for training are listed in Table~\ref{tab:data_composition}.

\begin{table}[!htpb]\centering\footnotesize
\renewcommand{\arraystretch}{1.0}
\caption{Training-set composition in training jets before multitask expansion. ``Other (each)'' is the quota assigned to each of the eight categories other than \texttt{Zqq} and \texttt{Wqq}.}
\label{tab:data_composition}
\begin{tabular}{l>{\columncolor[HTML]{CBCEFB}}rrrr}
\hline \hline
\textbf{Training Subset}   & \textbf{Total} & \texttt{Zqq} & \texttt{Wqq} & \textbf{Other (each)} \\
\hline
\Tsmall{}                  & 50,000  & 5,000  & 5,000  & 5,000  \\
\Tmedium{}                 & 100,000 & 10,000 & 10,000 & 10,000 \\
\Tlarge{}                  & 200,000 & 20,000 & 20,000 & 20,000 \\
\Tlarge{}-\WZthirty{}      & 200,000 & 30,000 & 30,000 & 17,500 \\
\Tlarge{}-\WZforty{}       & 200,000 & 40,000 & 40,000 & 15,000 \\
\hline \hline
\end{tabular}
\end{table}

\subsection{Four-panel jet representation}\label{sec:four_panel_representation}
\begin{figure}[!htpb]\centering
  \subfigure[]{
  \begin{minipage}[c]{0.40\linewidth}
  \includegraphics[width=1.0\linewidth]{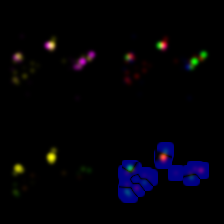}
  \end{minipage}}
  \subfigure[]{
  \begin{minipage}[c]{0.46\linewidth}
  \includegraphics[width=1.0\linewidth]{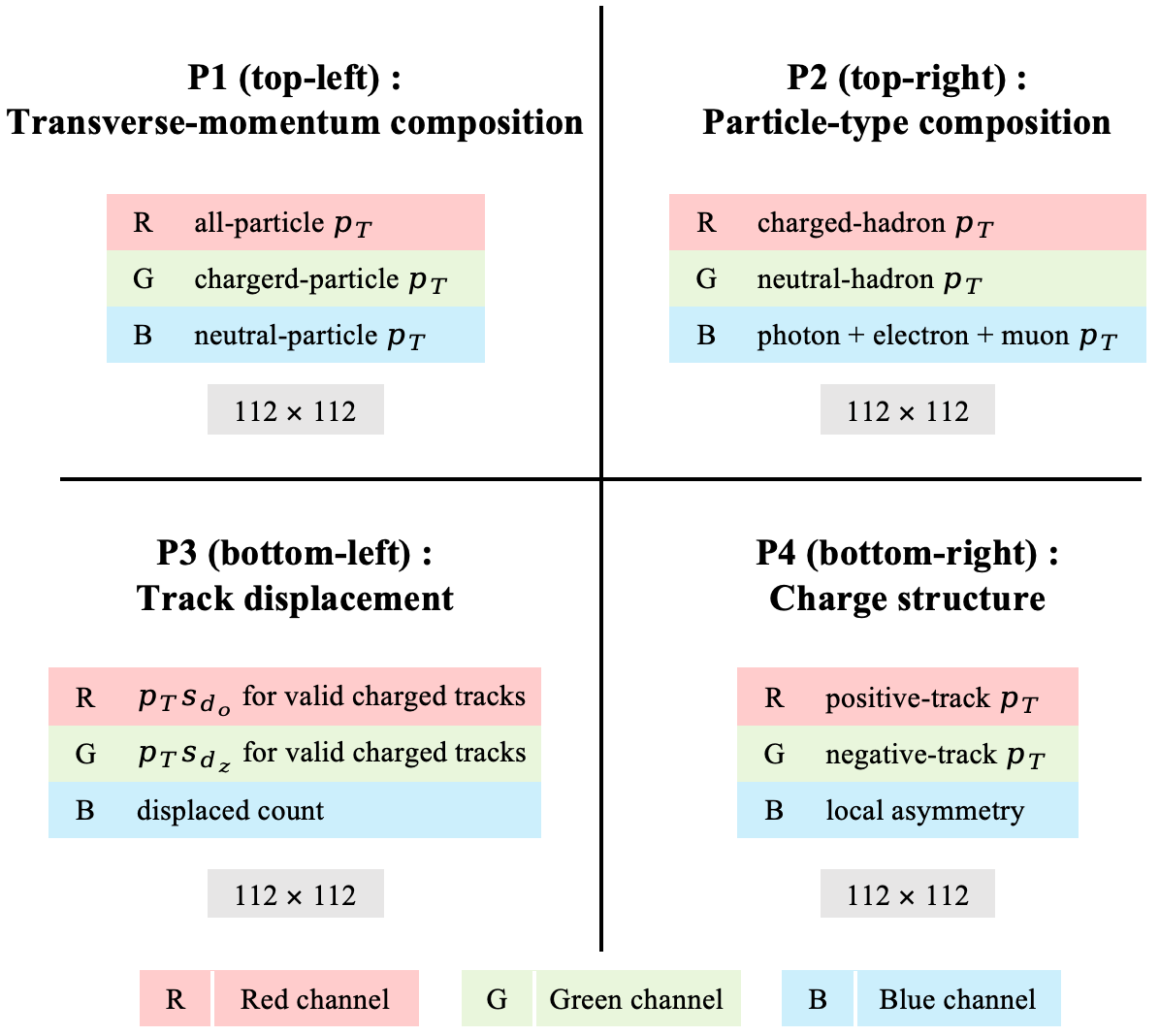}
  \end{minipage}}
  \caption{Four-panel jet representation and RGB-channel definitions. (a) The actual $224\times224$ model input for one validation jet. The RGB values represent the jointly normalized, logarithmically transformed channel intensities within each panel; absolute intensities are not comparable across panels or jets. (b) An explanatory schematic of the physical fields encoded in the four spatially aligned panels, each with a resolution of $112\times112$.}\label{fig:four_panel}
\end{figure}

Each model input pairs a plain-language instruction with one \(224\times224\) RGB image. No auxiliary numerical features or truth metadata are supplied to the model. The image is formed by arranging four \(112\times112\) RGB panels in a \(2\times2\) grid, with the RGB channels used to store three physical scalar fields rather than photographic color information. Before rasterization, each jet is geometrically canonicalized by centering its constituents, rotating the jet to align its principal axis when the covariance is sufficiently anisotropic, and applying deterministic reflections to resolve the remaining directional ambiguities. The same transformed coordinates are used for all twelve channels, thereby preserving pixel-wise spatial correspondence. Fig.~\ref{fig:four_panel}(a) shows the unannotated raster supplied to the model, whereas Fig.~\ref{fig:four_panel}(b) summarizes the channel definitions. The text, borders, and color keys in (b) are explanatory only and are not part of the model input. Unless otherwise stated, this subsection defines the complete preprocessing configuration used in the main experiments; variants with selected operations disabled are evaluated in Section~\ref{sec:results_panel_ablation}.

\paragraph{P1 (transverse-momentum composition) and P2 (particle species):}
For each jet, let \(i\) index its constituents, with transverse momentum \({\pt}_i=\sqrt{p_{x,i}^{2}+p_{y,i}^{2}}\), where \(p_{x,i}\) and \(p_{y,i}\) are the transverse momentum components of constituent \(i\). In P1, the red, green, and blue channels deposit the \(\pt\) of all, charged, and neutral particles, respectively. In P2, the corresponding channels contain the \(\pt\) densities of charged hadrons, neutral hadrons, and the combined photon--electron--muon category.

\paragraph{P3 (track displacement):}
For constituents identified as charged tracks, the clipped transverse and longitudinal displacement significances are defined by
\begin{equation}\label{eq:significance}
s_{d_0,i} = \min\!\left( \frac{|d_{0,i}|}{\max(\sigma_{d_0,i}, \epsilon_d)}, 5 \right),
\qquad
s_{d_z,i} = \min\!\left( \frac{|d_{z,i}|}{\max(\sigma_{d_z,i}, \epsilon_d)}, 5 \right)
\end{equation}
where \(d_{0,i}\) and \(d_{z,i}\) are the transverse and longitudinal impact parameters of track \(i\), while \(\sigma_{d_0,i}\) and \(\sigma_{d_z,i}\) are their corresponding uncertainties. The numerical floor is \(\epsilon_d=10^{-6}\). The absolute values remove the impact-parameter signs, while the upper bound of \(5\) limits the influence of extreme significances on the rendered intensity range.

A significance is valid only when the corresponding displacement and uncertainty are finite, and the uncertainty is strictly positive; otherwise, it is set to zero. The \(d_0\) and \(d_z\) validity conditions are evaluated independently. P3-R and P3-G deposit \({\pt}_i s_{d_0,i}\) and \({\pt}_i s_{d_z,i}\), respectively. P3-B deposits unit weight for charged tracks with at least one valid significance and \(\max(s_{d_0,i}, s_{d_z,i}) \geq 2\). Thus, \(5\) is a clipping bound, whereas \(2\) is the displaced-track threshold used by P3-B.

\paragraph{P4 (charge structure):}
P4-R and P4-G contain the \(\pt\) densities of positively and negatively charged tracks, respectively. To construct P4-B, intermediate positive- and negative-charge fields are formed using the conventional \(\pt^\kappa\) weighting employed in jet-charge observables~\cite{krohn2013jet}, with \(\kappa=0.5\):
\begin{equation}\label{eq:charge_fields}
P(u, v) = \sum_{i:q_i>0} p_{\mathrm T,i}^{\kappa} K_{\sigma}(u-u_i, v-v_i), 
\qquad
N(u, v) = \sum_{i:q_i<0} p_{\mathrm T,i}^{\kappa} K_{\sigma}(u-u_i, v-v_i)
\end{equation}
where \(q_i\) is the electric charge of constituent \(i\); \((u_i,v_i)\) is its transformed \((\deta,\dphi)\) position; \((u,v)\) denotes a location on the image grid; \(K_\sigma\) is the Gaussian deposition kernel defined below; and \(\kappa=0.5\) controls the transverse-momentum weighting.

The local absolute charge-field asymmetry is defined as
\begin{equation}\label{eq:charge_asymmetry}
A_{\mathrm{ch}}(u, v) = \frac{|P(u, v)-N(u, v)|}{P(u, v)+N(u, v) + \epsilon_{\mathrm{ch}}}
\end{equation}
where \(\epsilon_{\mathrm{ch}}=10^{-6}\) prevents division by zero in empty image regions. The P4-B field is defined as \(D_{\mathrm{P4},B}=A_{\mathrm{ch}}\). It measures the magnitude of the local imbalance, whereas P4-R and P4-G separately preserve the positive and negative contributions.

\paragraph{Geometric canonicalization:}
Following established jet-image canonicalization procedures~\cite{cogan2015jet,de2016jet}, deterministic, truth-label-independent geometric preprocessing is applied once to each jet, and the resulting coordinates are shared by all twelve channels. Constituents with non-finite \((\deta,\dphi,\pt)\) or non-positive \(\pt\) are removed. The remaining coordinates are translated to their \(\pt\)-weighted centroid, after which the residual \(\dphi\) values are wrapped to \([-\pi,\pi]\).

For PCA alignment, let \(\lambda_1\geq\lambda_2\geq0\) denote the eigenvalues of the \(\pt\)-weighted covariance matrix in the \((\dphi,\deta)\) plane. The covariance anisotropy is defined as
\begin{equation}\label{eq:pca_anisotropy}
a_{\mathrm{PCA}} = \frac{\lambda_1 - \lambda_2}{\lambda_1 + \lambda_2}
\end{equation}
When \(a_{\mathrm{PCA}}\geq0.05\), the coordinates are rotated so that the principal axis lies along the \(\dphi\) direction. Finally, the directional ambiguities are resolved through deterministic reflections: the leading-\(\pt\) constituent is placed in the half-plane \(\dphi\geq0\), and the non-leading transverse momentum is placed predominantly in the half-plane \(\deta<0\).

\paragraph{Rasterization and intensity encoding:}
Each directly deposited channel is rasterized on a uniform \(112\times112\) grid covering \(\deta,\dphi\in[-0.8,0.8]\). Let \(p\in\{\mathrm{P1},\mathrm{P2},\mathrm{P3},\mathrm{P4}\}\) denote the panel and \(c\in\{R,G,B\}\) its RGB channel. The directly deposited field is
\begin{equation}\label{eq:raster}
D_{p,c}(u,v) = \sum_{i\in S_{p,c}} w_{i,p,c} K_{\sigma}(u-u_i,v-v_i)
\end{equation}
where \(u\) and \(v\) are the \(\deta\) and \(\dphi\) coordinates of a pixel center, respectively; \(S_{p,c}\) is the constituent set selected for the channel; and \(w_{i,p,c}\geq0\) is the corresponding constituent weight.

Gaussian deposition is adopted to reduce sparse pixel occupancy, following related Gaussian-smearing treatments of jet images~\cite{dillon2023normalized, buss2023anomalous, oleksiyuk2024cluster}. The Gaussian kernel has width \(\sigma=0.03\) in \((\deta,\dphi)\) coordinate units, is truncated at \(4\sigma\), and is renormalized over the retained in-window pixels. Consequently, the deposited sum of each in-window constituent remains \(w_{i,p,c}\), including near an image boundary. For P3-B, \(w_{i,\mathrm{P3},B}=1\) for tracks passing the displacement threshold. P4-B is not deposited directly but is constructed from Eq.~\eqref{eq:charge_asymmetry}.

Following logarithmic intensity mappings used in multichannel jet images~\cite{komiske2017deep}, each direct or derived field is compressed as
\begin{equation}\label{eq:log_transform}
M_{p,c}(u,v) = \log\!\left[1+D_{p,c}(u,v)\right]
\end{equation}

The three channels in each panel are then jointly normalized:
\begin{equation}\label{eq:normalization}
m_p = \max_{c\in\{R, G, B\}, u, v}M_{p, c}(u, v),
\qquad
I_{p, c}(u, v) = \operatorname{round}\!\left[ 255\,\frac{M_{p, c}(u, v)}{m_p}\right]
\end{equation}
Here, \(m_p\) is the maximum transformed intensity in panel \(p\), and \(I_{p,c}\) is the stored 8-bit intensity of channel \(c\). All three channels are set to zero when \(m_p=0\). This normalization preserves relative spatial and channel intensities within each logarithmically transformed panel but removes absolute intensity differences across panels and jets. No stochastic image augmentation is applied.

\section{Instruction Tasks and Target Construction}\label{sec:tasks}
Building on the four-panel representation introduced in Section~\ref{sec:four_panel_representation}, three instruction tasks are defined.
Each example contains one image and a task-specific instruction, and the target is a compact \JSON{} object with a fixed schema and no surrounding prose.

\begin{figure}[!htpb]\centering
\includegraphics[width=0.78\linewidth]{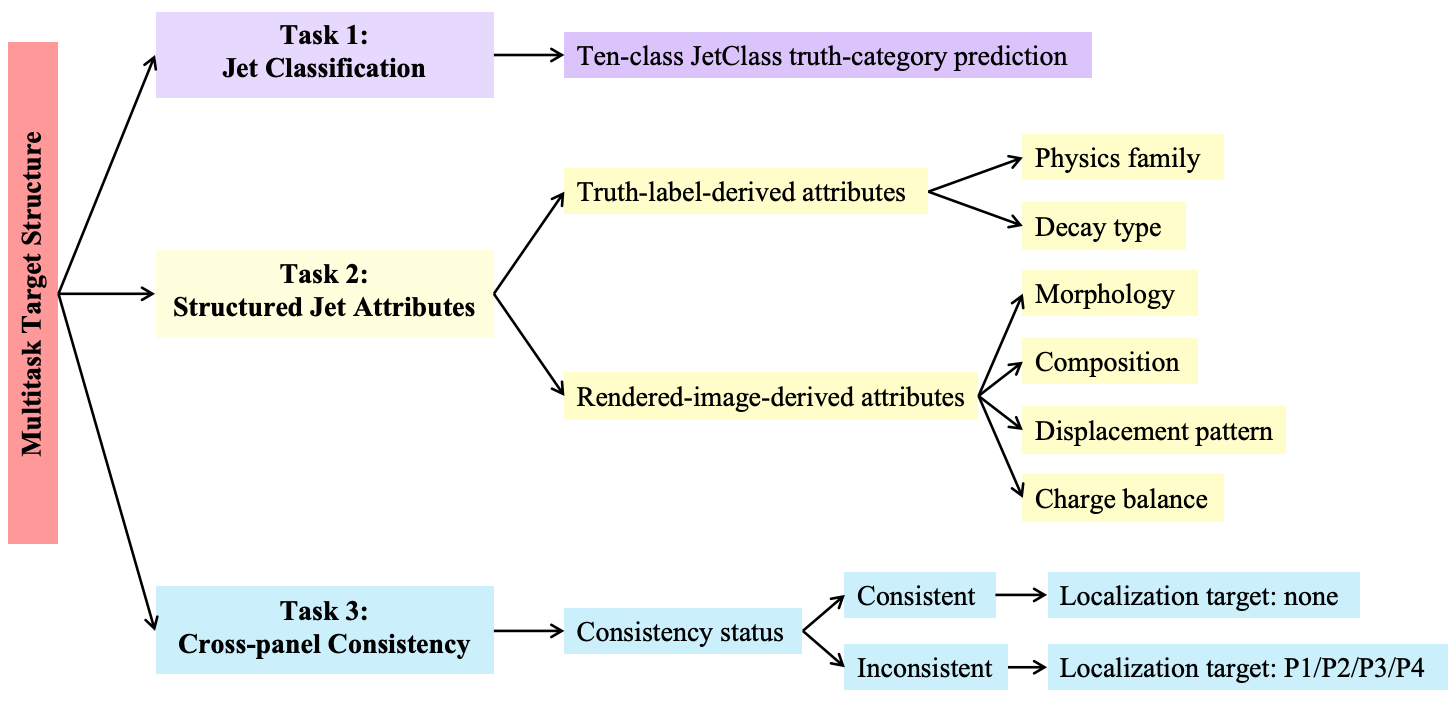}
\caption{Overview of the three instruction tasks and their target hierarchy. Task~1 predicts one of the ten JetClass truth categories listed in Table~\ref{tab:classes}. Task~2 comprises two attributes derived deterministically from the truth label and four attributes constructed from the rendered model input. Task~3 predicts the cross-panel consistency status and a conditional localization target: consistent examples use \texttt{none}, whereas inconsistent examples identify one of P1--P4 as the replaced panel. Human-readable labels are shown in the diagram; the exact \JSON{} field names, categorical vocabularies, and instruction templates are provided in Appendix~\ref{app:prompts}.}
\label{fig:overview_multitask}
\end{figure}

Fig.~\ref{fig:overview_multitask} summarizes the semantic structure of the three targets, which contain nine output fields in total: one for Task~1, six for Task~2, and two for Task~3. Appendix~\ref{app:prompts} gives the exact user instructions and output vocabularies. 

\paragraph{Task~1: Truth-category prediction.}
Task~1 predicts one of the ten JetClass categories in Table~\ref{tab:classes}. Its one-field target is taken directly from the Monte Carlo category, for example \texttt{\{"label":"Hbb"\}}.

\paragraph{Task~2: Structured attribute prediction.}
Task~2 predicts six categorical attributes. The \texttt{family} and \texttt{decay\_type} fields are deterministic coarsenings of the Task~1 label and therefore provide hierarchical class supervision rather than independent annotations. The other four fields are deterministic functions of the rendered RGB image supplied to the model; they are not generator-level variables or human annotations. Quantile thresholds are estimated from the corresponding training images and then applied unchanged to the held-out validation and test images. Appendix~\ref{app:task2_targets} specifies the mappings and image-based rules.

\begin{figure}[!htpb]\centering
\includegraphics[width=0.80\linewidth]{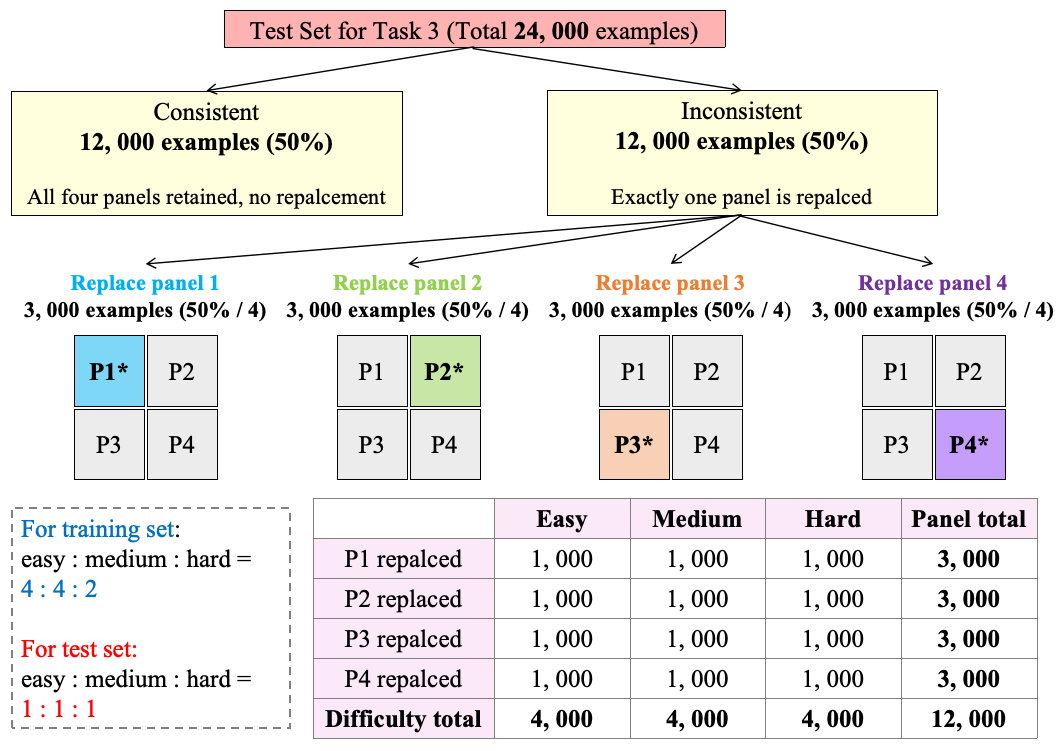}
\caption{Construction of the Task~3 evaluation examples in \test{}. For Task~3, the 24,000 examples are divided equally into 12,000 consistent and 12,000 inconsistent examples. The inconsistent half is jointly balanced over the four replacement locations and three donor strata, yielding 3,000 examples per panel, 4,000 per difficulty, and 1,000 per difficulty--panel combination. Colored panels marked with an asterisk indicate the replacement location. Training and validation mismatches follow a 4:4:2 difficulty mixture, whereas test mismatches use an equal 1:1:1 mixture.}
\label{fig:overview_testset}
\end{figure}

\paragraph{Task~3: Panel-replacement detection.}
Task~3 tests whether all four panels originate from the same source jet. Half of the examples are unchanged. In each remaining example, one $112\times112$ panel is replaced by the corresponding panel of a distinct donor jet from the same official split. The four replacement locations are balanced. Training and validation mismatches follow a 4:4:2 mixture of easy, medium, and hard donor strata, whereas test mismatches use an equal 1:1:1 mixture. Fig.~\ref{fig:overview_testset} summarizes the test-set composition, and Appendix~\ref{app:task3_construction} defines the strata and deterministic donor-selection procedure.
\section{Experimental Setup}\label{sec:setup}
\subsection{Models and training configuration}\label{sec:training_configuration}
\begin{figure}[!htpb]\centering
\includegraphics[width=0.78\linewidth]{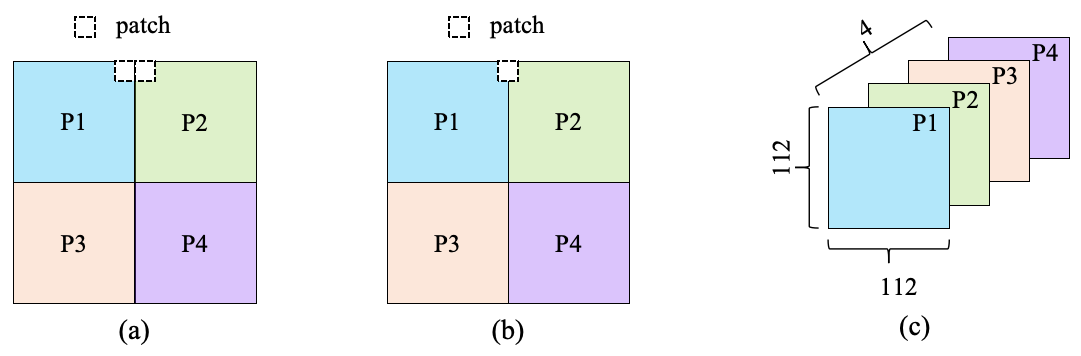}
\caption{Input organization and panel--patch alignment. (a) The configuration used in this study: four RGB panels are spatially tiled into one \(224\times224\times3\) image, with panel boundaries aligned to the visual patch grid. (b) A boundary-straddling patch, which the selected geometries avoid. (c) Four independently encoded \(112\times112\times3\) panel images, which are not used.}
\label{fig:patch_alignment}
\end{figure}

Four public vision--language checkpoints are adapted using LoRA~\cite{hu2022lora}:
\begin{itemize}
  \item \texttt{Qwen/Qwen2.5-VL-3B-Instruct} (\QwenThree{})~\cite{bai2025qwen25vl};
  \item \texttt{Qwen/Qwen2.5-VL-7B-Instruct} (\QwenSeven{})~\cite{bai2025qwen25vl};
  \item \texttt{google/paligemma2-3b-pt-224} (\PaliGemma{})~\cite{steiner2024paligemma};
  \item \texttt{google/gemma-4-E4B-it} (\GemmaFour{})~\cite{team2026gemma}.
\end{itemize}
They are selected to provide reproducible model access and complementary comparisons: the Qwen2.5-VL 3B and 7B checkpoints enable a within-family scale comparison, while \PaliGemma{} and \GemmaFour{} extend the evaluation across different multimodal architectures and pretraining recipes. Their visual-tokenization geometries are also compatible with the proposed \(2\times2\) layout. As illustrated in Fig.~\ref{fig:patch_alignment}(a), after any model-specific resizing, the panel boundaries coincide with both the visual patch grid and any local merging or pooling grid. Consequently, neither an initial patch nor a local aggregation region straddles two panels, avoiding the configuration illustrated in Fig.~\ref{fig:patch_alignment}(b). Each model receives one \(224\times224\times3\) RGB composite rather than four independently encoded \(112\times112\times3\) images, as contrasted in Fig.~\ref{fig:patch_alignment}(c). This design preserves the spatial organization of the four panels while allowing subsequent model layers to integrate information across them.

Table~\ref{tab:shared_training_config} lists the settings shared across models, and Table~\ref{tab:model_training_config} gives the model-dependent configuration. Cross-model results compare complete adaptation recipes rather than isolate any one architectural choice. Deterministic task-aware parsers require schema-valid \JSON{} for Tasks~2 and 3; for Task~1, they accept either valid \JSON{} or one unique canonical label.
An end-to-end example using the released inference utility is provided in Appendix~\ref{app:demo}.

\begin{table}[!htpb]\centering\footnotesize
\renewcommand{\arraystretch}{1.0}
\caption{Training settings shared by the adapted models.}
\label{tab:shared_training_config}
\begin{tabular}{ll}
\hline \hline
\textbf{Setting}            & \textbf{Value}                                                            
\\ \hline
Adaptation method           & LoRA; original checkpoint weights frozen                                      \\
Adapted components          & Language backbone, vision tower, and multimodal merger/projector              \\
LoRA rank $r$               & 32                                                                            \\
LoRA scaling $\alpha$       & 64                                                                            \\
LoRA dropout                & 0.05                                                                          \\
Training epochs             & 3                                                                             \\
Optimizer                   & Fused PyTorch AdamW (\texttt{adamw\_torch\_fused})                            \\
Optimizer hyperparameters   & $\beta_1=0.9$,  $\beta_2=0.999$,  $\epsilon_{\mathrm{Adam}}=10^{-8}$          \\
Peak learning rate          & $8\times10^{-5}$                                                              \\
Learning-rate schedule      & Linear warm-up followed by cosine decay                                       \\
Weight decay                & 0                                                                             \\
Maximum gradient norm       & 1.0                                                                           \\
Compute precision           & bfloat16                                                                      \\
Supervised-loss mask        & Assistant/completion target tokens only                                       \\
Training hardware           & \(2\times\) NVIDIA H200 GPUs (143,771 MiB each)                               \\
Software environment        & PyTorch 2.8.0 (CUDA 12.8 build)                                               \\ \hline \hline
\end{tabular}
\end{table}

\begin{table}[!htpb]\centering\footnotesize
\renewcommand{\arraystretch}{1.0}
\caption{Model-dependent configuration for one rendered $224\times224$ four-panel input. Visual tokens are the image embeddings supplied to the language model; resolved LoRA modules are the matched module instances across all adapted components.}
\label{tab:model_training_config}
\begin{tabular}{lllll}
\hline \hline
\textbf{Setting}  & \textbf{\QwenThree{}}  & \textbf{\QwenSeven{}}  & \textbf{\PaliGemma{}}  & \textbf{\GemmaFour{}}  \\ \hline
Rendered input  & $224\times224$  & $224\times224$  & $224\times224$  & $224\times224$  \\ 
Processor image size  & $224\times224$  & $224\times224$  & $224\times224$  & $768\times768$  \\
Vision patch size  & $14\times14$  & $14\times14$  & $14\times14$  & $16\times16$ \\
Pre-merge/pooling grid  & $16\times16$  & $16\times16$  & $16\times16$  & $48\times48$  \\
Vision resampling  & $2\times2$ merge & $2\times2$ merge & None & $3\times3$ average pooling \\
Output visual grid  & $8\times8$  & $8\times8$  & $16\times16$  & $16\times16$ \\
Visual tokens  & 64  & 64  & 256  & 256  \\
Resolved LoRA modules  & 414  & 358  & 345  & 371 \\
Trainable LoRA parameters  & 82.73M  & 103.65M  & 59.03M  & 78.91M \\
Per-GPU batch  & 256  & 256  & 64  & 64 \\
Effective global batch  & 512  & 512  & 128  & 128 \\
Maximum sequence length  & 512  & 512  & 512  & 1024 \\
Warm-up  & 50 steps  & 50 steps  & 5\% of optimizer steps & 5\% of optimizer steps \\
Training attention  & FlashAttention-2  & FlashAttention-2  & SDPA  & SDPA \\
Inference backend  & vLLM  & vLLM  & Hugging Face  & Hugging Face \\
\hline \hline
\end{tabular}
\end{table}

\subsection{Experimental comparisons and controls}\label{sec:experimental_comparisons}
Table~\ref{tab:study_matrix} organizes the experiments according to the factor examined in each comparison. The preprocessing ablation varies orientation canonicalization, Gaussian deposition, and logarithmic intensity compression. The panel ablation changes the visible image content; the multitask study changes the training-task schedule; the zero-shot control evaluates the unadapted checkpoints; the balanced-scaling study changes the number of training source jets; and the \(W/Z\)-enrichment study redistributes a fixed 200,000-jet budget among classes. The cross-dataset transfer study varies the target data domain and adaptation strategy. Within each study, the model, source-jet selection, evaluation set, and training configuration are held fixed whenever applicable, except for the factor explicitly under investigation.

\begin{table}[!htpb]\centering\footnotesize
\renewcommand{\arraystretch}{1.0}
\caption{Experimental comparisons. Dataset names denote the number of distinct source jets before multitask expansion.}
\label{tab:study_matrix}
\begin{tabular}{llll}
\hline \hline
\textbf{Study}     & \textbf{Model(s)}         & \textbf{Data or configurations}        & \textbf{Protocol}  
\\ \hline
Preprocessing ablation & \PaliGemma{} & \Tsmall{} & Task~1 only  \\
\rowcolor[HTML]{EFEFEF}
Panel ablation     & \PaliGemma{}              & \Tsmall{}                              & Task~1 only    \\ 
Multitask effect   & \PaliGemma{}              & \Tsmall{}                              & Task~1 only vs.\ Tasks~1--3 \\ 
\rowcolor[HTML]{EFEFEF}
Zero-shot control  & All four base checkpoints   & No training data                       & Task~1 only    \\
Balanced scaling   & All four models   & \Tsmall{}, \Tmedium{}, and \Tlarge{}   & Tasks~1--3     \\ 
\rowcolor[HTML]{EFEFEF}
\(W/Z\) enrichment 
& \begin{tabular}[c]{@{}l@{}}\QwenThree{}, \\ \QwenSeven{}, \\ and \PaliGemma{}\end{tabular} 
& \begin{tabular}[c]{@{}l@{}}\Tlarge{}, \\ \Tlarge{}-\WZthirty{}, \\ and \Tlarge{}-\WZforty{}\end{tabular} 
& Tasks~1--3  \\ 
Cross-dataset transfer 
& \begin{tabular}[c]{@{}l@{}}\PaliGemma{}, \\ and \GemmaFour{}\end{tabular} 
& \begin{tabular}[c]{@{}l@{}}\Tlarge{} adapter, \\JetClass-II, \\Aspen Open Jets\end{tabular}
& Tasks~1--3  \\ 
\hline \hline
\end{tabular}
\end{table}

\paragraph{$\ast$ Preprocessing ablation:}
This experiment assesses the effects of orientation canonicalization, Gaussian deposition, and logarithmic intensity compression. Five Task~1-only \PaliGemma{} adapters are trained on \Tsmall{} using the complete preprocessing pipeline, with each component disabled individually, or with all three components disabled. The variants are compared only on the fixed V5k validation subset, and the complete preprocessing configuration is frozen before evaluation on \test{}.
  
\paragraph{$\ast$ Panel ablation:}
This experiment quantifies the individual and combined contributions of particle species, displacement, and charge information to Task~1 classification. Five Task~1-only \PaliGemma{} variants are trained on \Tsmall{}: P1, P1+P2, P1+P2+P3, P1+P2+P4, and P1+P2+P3+P4. Panels omitted from each variant are replaced with zero-valued pixels without resizing or rearranging the \(224\times224\) canvas. 

\paragraph{$\ast$ Multitask effect:}
This comparison tests how joint supervision from Tasks~2 and 3 affects Task~1 classification while adding the two auxiliary capabilities. The full-panel (P1+P2+P3+P4) Task~1-only \PaliGemma{} model at \Tsmall{} is compared with the corresponding three-task model. The two runs use the same source jets, but their per-epoch schedules contain 50,000 and 150,000 task records, respectively. At the same epoch count and global batch, the multitask run therefore performs approximately three times as many optimizer updates. Accordingly, this comparison evaluates complete training recipes rather than auxiliary-task transfer at fixed compute.

\paragraph{$\ast$ Zero-shot control:}
This control tests whether generic VLM pretraining is sufficient to interpret the four-panel jet representation without domain-specific adaptation. The label-and-schema prompted zero-shot lists the ten allowed labels and the required one-field \JSON{} schema, which is shown in Appendix~\ref{app:prompts}; a strict zero-shot uses the short adaptation prompt without listing the labels. The two prompting conditions are summarized in Table~\ref{tab:zero_shot_prompts}.

\begin{table}[!htpb]\centering\footnotesize
\renewcommand{\arraystretch}{1.0}
\caption{Zero-shot prompting conditions for Task~1. Neither condition uses task-specific parameter updates, in-context demonstrations, or class descriptions.}
\label{tab:zero_shot_prompts}
\begin{tabularx}{\linewidth}{@{}cYY@{}}
\hline \hline
\textbf{Setting} & \textbf{Information supplied} & \textbf{Role in the analysis} \\
\hline
\rowcolor[HTML]{EFEFEF}
Strict zero-shot
& The short Task~1 question used during adaptation, without the JetClass label vocabulary or the required output schema.
& Diagnostic of whether an unadapted checkpoint can recognize the task vocabulary and response format without explicit constraints. \\ 
Label-and-schema prompted zero-shot
& The ten allowed labels and the required one-field \JSON{} schema, without demonstrations, class descriptions, or jet-physics guidance.
& Practical zero-training reference, with output-interface ambiguity reduced.
\\ \hline \hline
\end{tabularx}
\end{table}

\paragraph{$\ast$ Balanced scaling:}
This study measures how performance on the three tasks changes with the amount of balanced training data across the four model recipes. All four models are adapted on the class-balanced \Tsmall{}, \Tmedium{}, and \Tlarge{} subsets. 

\paragraph{$\ast$ Fixed-budget \(W/Z\) enrichment:}
This experiment tests whether allocating more training examples to the \texttt{Zqq} and \texttt{Wqq} jet categories improves their recognition at fixed total data size. At \Tlarge{}, \QwenThree{}, \QwenSeven{}, and \PaliGemma{} are trained with the balanced, \WZthirty{}, and \WZforty{} allocations defined in Table~\ref{tab:data_composition}. The total training budget remains 200,000 source jets. The design therefore measures data reallocation, not the effect of adding \texttt{Zqq} and \texttt{Wqq} examples without reducing the other class quotas.

\paragraph{$\ast$ Cross-dataset transfer:}
This experiment evaluates whether the JetClass-trained adapters generalize to other datasets without changing the four-panel representation. It compares random baselines, frozen adapter transfer, and limited target-domain adaptation.

\subsection{Evaluation metrics}\label{sec:evaluation_metrics}
To quantify the results reported in Section~\ref{sec:results}, the evaluation metrics are defined as follows~\cite{grosso2026veraiphy, sokolova2009systematic}. For each class \(j\): 
\begin{itemize}
    \item A true positive (\(\mathrm{TP}_j\)) is an example whose true and predicted labels are both \(j\).
    \item A false positive (\(\mathrm{FP}_j\)) is predicted as \(j\) but belongs to another class.
    \item A false negative (\(\mathrm{FN}_j\)) belongs to \(j\) but is predicted as another class.
    \item A true negative (\(\mathrm{TN}_j\)) neither belongs to nor is predicted as \(j\).
\end{itemize}

The precision, recall, and \(F_1\) score for class \(j\) are:
\begin{equation}
\mathrm{Precision}_j = \frac{\mathrm{TP}_j}{\mathrm{TP}_j + \mathrm{FP}_j},
\qquad
\mathrm{Recall}_j = \frac{\mathrm{TP}_j}{\mathrm{TP}_j + \mathrm{FN}_j},
\qquad
F_{1,j} = \frac{2 \times \mathrm{Precision}_j \times \mathrm{Recall}_j}{\mathrm{Precision}_j + \mathrm{Recall}_j}
\end{equation}
  
For \(J\) classes, the Macro metrics are the unweighted averages:
\begin{equation}
\mathrm{Precision}_{\mathrm{macro}} = \frac{1}{J}\sum_{j=1}^{J} \mathrm{Precision}_j,
\qquad
\mathrm{Recall}_{\mathrm{macro}} = \frac{1}{J}\sum_{j=1}^{J} \mathrm{Recall}_j,
\qquad
F_{1,\mathrm{macro}} = \frac{1}{J}\sum_{j=1}^{J} F_{1,j}
\end{equation}

Overall accuracy is the fraction of correctly classified examples:
\begin{equation}
\mathrm{Accuracy} = \frac{1}{n_{\mathrm{tot}}}\sum_{j=1}^{J}\mathrm{TP}_j
\end{equation}
where \(n_{\mathrm{tot}}\) is the total number of examples in the dataset being evaluated.

Because the V5k validation subset and \test{} test subset contain equal numbers of jets from each Task~1 category, Task~1 overall accuracy and Macro recall are numerically identical. 
For Task~2, recall and \(F_1\) are computed separately for each allowed categorical value of every output field. The per-field Macro recall and Macro \(F_1\) are obtained by unweighted averaging over the allowed values. Field-mean Macro recall and field-mean Macro \(F_1\) are then computed by averaging the corresponding per-field metrics equally over the six fields. 
For Task~3, binary consistency performance is reported using the recalls of the consistent and inconsistent classes together with their Macro recall. Replaced-panel localization is evaluated only on inconsistent examples; recall is computed separately for P1--P4, and their unweighted mean is reported as the localization Macro recall.
\section{Results}\label{sec:results}
\subsection{Preprocessing pipeline and systematic variations}\label{sec:results_panel_ablation}
Before assessing the information contributed by individual panels, three preprocessing components are examined using five independently trained Task~1-only \PaliGemma{} adapters at \Tsmall{}, without using \test{}: orientation canonicalization, Gaussian deposition, and logarithmic intensity compression. Here, orientation canonicalization denotes conditional PCA rotation followed by deterministic reflections; \(\pt\)-weighted centering is retained in every variant. Three variants disable one component at a time, while the Minimal variant disables all three.

Table~\ref{tab:preprocessing_ablation} shows the results. Active pixels measures the mean fraction of image positions containing at least one nonzero RGB value. Orientation robustness is the fraction of predictions on seven rotated or reflected views that match the prediction for the corresponding original image. This diagnostic is evaluated only for the Full and No orientation canonicalization variants to isolate the effect of orientation canonicalization. The results show that Gaussian deposition increases the active-pixel fraction from \(0.21\%\) to \(10.42\%\). Removing either Gaussian deposition or logarithmic intensity compression reduces Macro recall by \(5.02\) percentage points. Orientation canonicalization produces a smaller gain of \(0.48\) percentage points in Macro recall, but increases orientation robustness from \(84.36\%\) to \(99.60\%\). It therefore primarily suppresses nuisance dependence on jet orientation, while Gaussian deposition and logarithmic compression improve the accessibility of sparse and weak activity. The complete preprocessing configuration is fixed before evaluation on \test{}.

\begin{table}[!htpb]\centering\footnotesize
\renewcommand{\arraystretch}{1.0}
\caption{Validation-only Task~1 preprocessing ablation for \PaliGemma{} trained on \Tsmall{}. Minimal disables all three varied components, while filtering, \(\pt\)-weighted centering, raster geometry, and per-panel normalization remain fixed. Orientation robustness is the mean agreement between predictions on seven rotated or reflected views and the corresponding original-view prediction. All numerical entries are percentages; bold marks the best result, and dashes denote unevaluated cases.}
\label{tab:preprocessing_ablation}
\begin{tabular}{ccccccccc}
\hline \hline
\multirow{2}{*}{Preprocessing} 
& \multirow{2}{*}{\begin{tabular}[c]{@{}c@{}}Orientation\\ canonicalization\end{tabular}} 
& \multirow{2}{*}{Gaussian} 
& \multirow{2}{*}{Log}   
& \multirow{2}{*}{\begin{tabular}[c]{@{}c@{}}Active\\ pixels\end{tabular}} 
& \multirow{2}{*}{\begin{tabular}[c]{@{}c@{}}Macro\\ recall\end{tabular}} 
& \multirow{2}{*}{\begin{tabular}[c]{@{}c@{}}Macro\\ precision\end{tabular}} 
& \multirow{2}{*}{\begin{tabular}[c]{@{}c@{}}Macro\\ \(F_1\)\end{tabular}} 
& \multirow{2}{*}{\begin{tabular}[c]{@{}c@{}}Orientation\\ robustness\end{tabular}} \\
\\ \hline 
Full     
& \ding{51} & \ding{51} & \ding{51} 
& 10.42 & \textbf{68.62} & \textbf{68.36}  & \textbf{68.20} & \textbf{99.60} \\
Orientation off         
& \ding{55} & \ding{51} & \ding{51} 
& 10.42 & 68.14 & 67.84 & 67.57 & 84.36 \\
Gaussian off         
& \ding{51} & \ding{55} & \ding{51} 
& 0.21  & 63.60  & 63.02  & 62.48  & --    \\
Log off      
& \ding{51} & \ding{51} & \ding{55} 
& 8.72  & 63.60  & 63.17  & 62.98  & --    \\
Minimal  
& \ding{55} & \ding{55} & \ding{55} 
& 0.21  & 59.30  & 59.09  & 57.83  & --   \\ 
\hline \hline
\end{tabular}
\end{table}

After fixing the preprocessing configuration using validation data only, the panel ablation examines the incremental contribution of each physical panel to Task~1 classification. Five \PaliGemma{} adapters are trained separately using P1, P1+P2, P1+P2+P3, P1+P2+P4, or the complete P1+P2+P3+P4 representation. All variants use \Tsmall{} for training. Removed panels are replaced by zero-valued pixels while the \(224\times224\) canvas and panel positions remain unchanged, thereby preserving the processor geometry and visual-token count. The five adapters are evaluated on the fixed \test{} benchmark.

Table~\ref{tab:panel_ablation_test24k} reports the per-category recall and Macro recall. The results show that: (i) Adding the particle-species panel P2 to P1 increases the Macro recall from \(57.50\%\) to \(61.29\%\). (ii) Adding the displacement-sensitive P3 panel to P1+P2 produces a larger improvement, reaching \(67.47\%\), whereas adding the charge-sensitive P4 panel instead gives \(62.82\%\). (iii) P4 nevertheless provides a pronounced improvement for \texttt{Zqq}, whose recall increases from \(22.54\%\) with P1+P2 to \(34.21\%\) with P1+P2+P4. (iv) The complete four-panel representation achieves the highest Macro recall, \(68.80\%\), corresponding to an absolute improvement of \(11.30\) percentage points over P1 alone. 
These results indicate that the panels encode complementary information: displacement is particularly important for the overall classification performance, while charge information provides additional discrimination for selected categories. Therefore, all four panels are kept for the following experiments.

\begin{table}[!htpb]\centering\footnotesize
\renewcommand{\arraystretch}{1.0}
\caption{Task~1 per-category and Macro recall (\%) for the \PaliGemma{} panel-ablation adapters trained on \Tsmall{} and evaluated on the same balanced \test{} benchmark. Bold denotes the best result in each column.}
\label{tab:panel_ablation_test24k}
\begin{tabular}{ccccccccccc
>{\columncolor[HTML]{CBCEFB}}c }
\hline \hline
                           & \multicolumn{10}{c}{Per-category recall (\%)}   & \cellcolor[HTML]{CBCEFB}   \\ \cline{2-11}
\multirow{-2}{*}{Panels} 
& \texttt{QCD} & \texttt{Hbb} & \texttt{Hcc} & \texttt{Hgg} & \texttt{H4q} & \texttt{Hqql} & \texttt{Zqq} & \texttt{Wqq} & \texttt{Tbqq} & \texttt{Tbl} 
& \multirow{-2}{*}{Macro recall}  \\ 
\hline
P1 & 62.54 & 29.46 & 40.50 & 58.63 & 70.29 & 76.21 & 19.46 & 55.08 & 76.25 & 86.54 & 57.50 \\
P1+P2 & 63.92 & 34.33 & 45.79 & 57.63 & 71.00 & 87.21 & 22.54 & 60.92 & 78.71 & 90.83 & 61.29 \\
P1+P2+P3 & 66.25 & \textbf{70.00} & 55.29 & \textbf{61.63} & 70.75 & \textbf{90.54} & 25.25 & 60.88 & 81.79 & \textbf{92.33} & 67.47 \\
P1+P2+P4 & 64.63 & 34.25 & 44.83 & 56.58 & \textbf{72.75} & 86.71 & \textbf{34.21} & 60.88 & 82.88 & 90.46 & 62.82 \\
P1+P2+P3+P4 & \textbf{67.54} & 69.04 & \textbf{55.79} & \textbf{61.63} & 71.75 & 90.04 & 33.83 & \textbf{63.04} & \textbf{83.38} & 92.00 & \textbf{68.80} \\
\hline \hline
\end{tabular}

\end{table}

\subsection{Multitask effect}
Having established the benefit of the complete four-panel representation, the analysis next examines whether auxiliary supervision from Tasks~2 and 3 affects the primary truth-classification task. Two full-panel \PaliGemma{} adapters initialized from the same checkpoint and trained on \Tsmall{} are compared. The Task~1-only schedule contains one classification record per source jet, whereas the multitask schedule contains one record for each of Tasks~1--3. Both adapters are evaluated exclusively on \test{}. 

Table~\ref{tab:multitask_task1_test24k} reports the per-category and aggregate classification metrics. The results show the following. (i) The multitask recipe increases the Task~1 Macro recall from \(68.80\%\) to \(70.21\%\), a gain of \(1.41\) percentage points, with corresponding gains of \(1.46\) and \(1.45\) percentage points in Macro precision and Macro \(F_1\). (ii) Per-category recall improves for all ten truth categories, with the largest gains for \texttt{Zqq}, \texttt{Hgg}, and \texttt{Hbb}, at \(3.25\), \(3.21\), and \(2.67\) percentage points, respectively. (iii) The improvement for \texttt{Zqq} is accompanied by gains of \(4.30\) percentage points in precision and \(3.71\) percentage points in \(F_1\). (iv) \texttt{QCD} is the only category with a lower precision, decreasing by \(0.97\) percentage points, although its higher recall leaves its \(F_1\) essentially unchanged. Overall, the auxiliary targets are compatible with the primary classification objective and yield a modest but consistent improvement in Task~1 performance under the complete multitask training recipe.

\begin{table}[!htpb]\centering\footnotesize
\renewcommand{\arraystretch}{1.0}
\caption{Task~1 classification metrics (\%) for full-panel \PaliGemma{} adapters trained on \Tsmall{} with Task~1 only (T1) or Tasks~1--3 (MT) and evaluated on \test{}. The Macro row gives the unweighted average over the ten categories. Differences \(\Delta=\mathrm{MT}-\mathrm{T1}\) are reported in percentage points and computed before rounding.}
\label{tab:multitask_task1_test24k}
\begin{tabular}{l rrr rrr rrr}
\hline \hline
\multirow{2}{*}{Category}
& \multicolumn{3}{c}{Recall (\%)}
& \multicolumn{3}{c}{Precision (\%)}
& \multicolumn{3}{c}{\(F_1\) (\%)} \\
\cmidrule(lr){2-4}
\cmidrule(lr){5-7}
\cmidrule(lr){8-10}

& T1 & MT & \(\Delta\)
& T1 & MT & \(\Delta\)
& T1 & MT & \(\Delta\) \\
\hline
\texttt{QCD}
& 67.54 & 68.38 & \(+0.83\)
& 75.29 & 74.32 & \cellcolor[HTML]{FFE2E0}\(-0.97\)
& 71.21 & 71.22 & \(+0.02\) \\
\texttt{Hbb}
& 69.04 & 71.71 & \(+2.67\)
& 70.42 & 70.45 & \(+0.03\)
& 69.72 & 71.07 & \(+1.35\) \\
\texttt{Hcc}
& 55.79 & 56.67 & \(+0.88\)
& 57.10 & 58.22 & \(+1.12\)
& 56.44 & 57.43 & \(+0.99\) \\
\texttt{Hgg}
& 61.63 & 64.83 & \(+3.21\)
& 58.16 & 58.43 & \(+0.27\)
& 59.84 & 61.47 & \(+1.62\) \\
\texttt{H4q}
& 71.75 & 72.54 & \(+0.79\)
& 60.19 & 62.67 & \(+2.48\)
& 65.46 & 67.25 & \(+1.78\) \\
\texttt{Hqql}
& 90.04 & 90.75 & \(+0.71\)
& 86.51 & 88.18 & \(+1.67\)
& 88.24 & 89.45 & \(+1.21\) \\
\texttt{Zqq}
& 33.83 & 37.08 & \(+3.25\)
& 50.43 & 54.74 & \(+4.30\)
& 40.50 & 44.21 & \(+3.71\) \\
\texttt{Wqq}
& 63.04 & 63.25 & \(+0.21\)
& 57.22 & 60.33 & \(+3.11\)
& 59.99 & 61.76 & \(+1.77\) \\
\texttt{Tbqq}
& 83.38 & 84.79 & \(+1.42\)
& 76.40 & 78.03 & \(+1.63\)
& 79.74 & 81.27 & \(+1.53\) \\
\texttt{Tbl}
& 92.00 & 92.13 & \(+0.13\)
& 93.01 & 94.01 & \(+1.00\)
& 92.50 & 93.06 & \(+0.55\) \\
\hline
\rowcolor[HTML]{CBCEFB}
\textbf{Macro}
& 68.80 & \textbf{70.21} & \(\mathbf{+1.41}\)
& 68.47 & \textbf{69.94} & \(\mathbf{+1.46}\)
& 68.36 & \textbf{69.82} & \(\mathbf{+1.45}\) \\
\hline \hline
\end{tabular}
\end{table}

\subsection{Zero-shot and balanced scaling}\label{sec:scaling_results}
\subsubsection{Zero-shot prompting and performance}
Each unadapted checkpoint is evaluated on \test{} under the two prompting conditions defined in Table~\ref{tab:zero_shot_prompts}. Table~\ref{tab:zero_shot_test24k} shows that strict zero-shot produces invalid-output rates of \(88.12\%--100\%\), indicating that the short prompt does not reliably specify the canonical JetClass vocabulary or response format. The label-and-schema prompt reduces the invalid-output rate to zero for both Qwen checkpoints and \GemmaFour{}, but their Macro recalls remain close to the \(10\%\) chance level, and their Macro \(F_1\) scores range from \(1.83\%\) to \(2.87\%\). \PaliGemma{} retains a \(64.94\%\) invalid-output rate and reaches only \(3.33\%\) Macro recall. Explicit label and schema information therefore reduces output-interface ambiguity but does not provide useful discrimination of the four-panel representation.

\begin{table}[!htpb]\centering\footnotesize
\renewcommand{\arraystretch}{1.0}
\caption{Zero-shot Task~1 performance on the frozen \test{} benchmark. No adapter is loaded. Invalid-output rate denotes the fraction of responses that cannot be parsed as one allowed Task~1 label; these outputs are counted as incorrect when computing Macro recall and Macro \(F_1\).}
\label{tab:zero_shot_test24k}
\begin{tabular}{@{}crrr@{\hspace{12pt}}rrr@{}}
\hline \hline
\multirow{2}{*}{Model}
&\multicolumn{3}{c}{Strict zero-shot}
&\multicolumn{3}{c}{\shortstack{Label-and-schema prompted zero-shot}}
\\ \cmidrule(lr){2-4} \cmidrule(lr){5-7}
& Macro recall (\%) & Macro \(F_1\) (\%)  & Invalid (\%)
& Macro recall (\%) & Macro \(F_1\) (\%)  & Invalid (\%)
\\ \hline
\QwenThree{} & 0.00 & 0.00 & 99.78 & 9.88  & 2.15 & 0.00  \\
\QwenSeven{} & 0.97 & 0.88 & 88.12 & 10.57 & 2.87 & 0.00  \\
\PaliGemma{} & 0.00 & 0.00 & 100.00 & 3.33 & 1.48 & 64.94 \\
\GemmaFour{} & 0.00 & 0.01 & 99.91 & 10.00 & 1.83 & 0.00  \\
\hline \hline
\end{tabular}

\end{table}

The label-and-schema prompted condition serves as the practical zero-training point in the balanced-data scaling study. Because its instruction is more explicit than the prompt used for adapted models, it is a prompt-assisted reference rather than a strictly prompt-matched point on the scaling curve.

\subsubsection{Scaling law for balanced training data}\label{sec:scalinglaw_balanced}
\begin{figure}[!htpb]\centering
\includegraphics[width=0.92\linewidth]{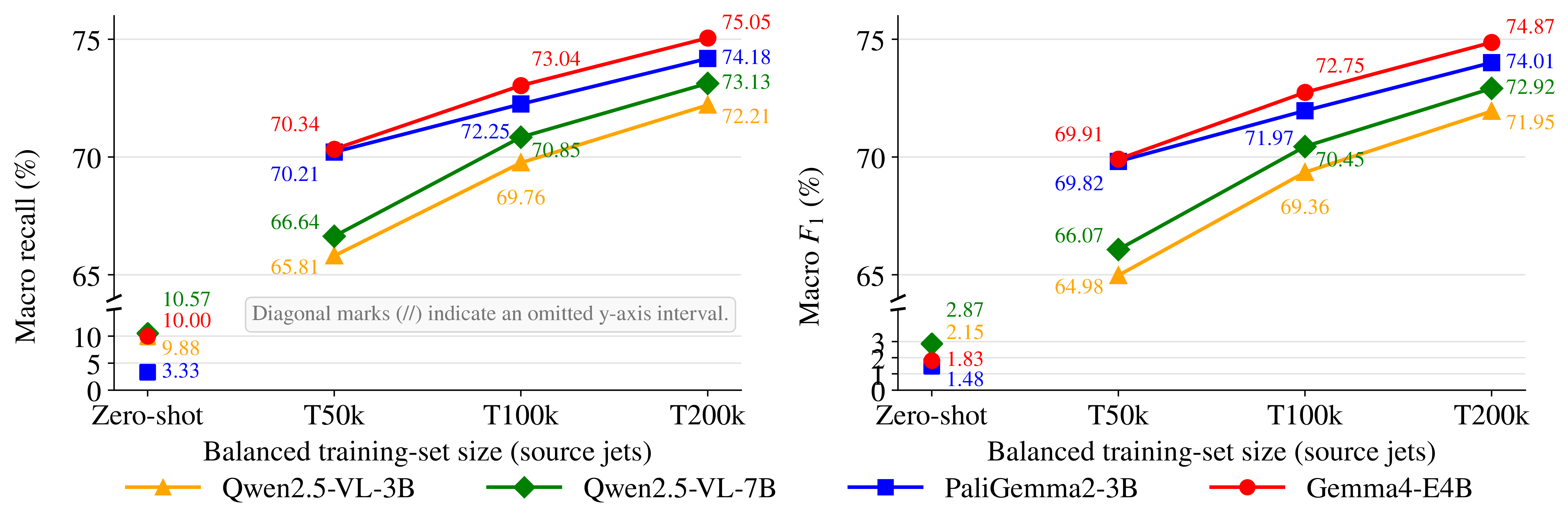}
\caption{Prompted zero-shot reference and balanced-data scaling of Task~1 performance. Macro recall (equivalent to overall accuracy on the class-balanced benchmark, left) and Macro \(F_1\) (right) are evaluated on the frozen \test{} benchmark for four models. Zero-shot denotes label-and-schema prompted zero-shot evaluation without an adapter, whereas \Tsmall{}, \Tmedium{}, and \Tlarge{} denote LoRA-adapted checkpoints trained on class-balanced samples containing 50,000, 100,000, and 200,000 source jets, respectively. The broken \(y\)-axes separate the zero-shot and adapted regimes.}
\label{fig:MT_task1}
\end{figure}

\paragraph{Task~1:}
Fig.~\ref{fig:MT_task1} shows a clear transition from near-chance zero-shot performance to effective Task~1 classification after domain adaptation. (i) Under the zero-shot condition, three checkpoints obtain Macro recall values close to the \(10\%\) chance level of the balanced ten-class benchmark, but their Macro \(F_1\) scores remain below \(3\%\). (ii) After adaptation on \Tsmall{}, Macro recall rises to \(65.81\%-70.34\%\), while Macro \(F_1\) reaches \(64.98\%-69.91\%\). Both metrics then improve monotonically with training-set size for every model. (iii) At \Tlarge{}, Macro recall spans \(72.21\%-75.05\%\), and Macro \(F_1\) spans \(71.95\%-74.87\%\). (iv) \GemmaFour{} gives the highest result at each adapted scale, followed closely by \PaliGemma{}, while \QwenSeven{} consistently outperforms \QwenThree{}. (v) The gains from \Tmedium{} to \Tlarge{} are smaller than those from \Tsmall{} to \Tmedium{} for all four checkpoints, indicating diminishing returns as the balanced training sample grows. However, performance continues to increase at \Tlarge{}, and the present results do not establish a clear saturation point. 

\begin{figure}[!htpb]\centering
  \subfigure[]{
  \begin{minipage}[c]{0.77\linewidth}
  \includegraphics[width=1.0\linewidth]{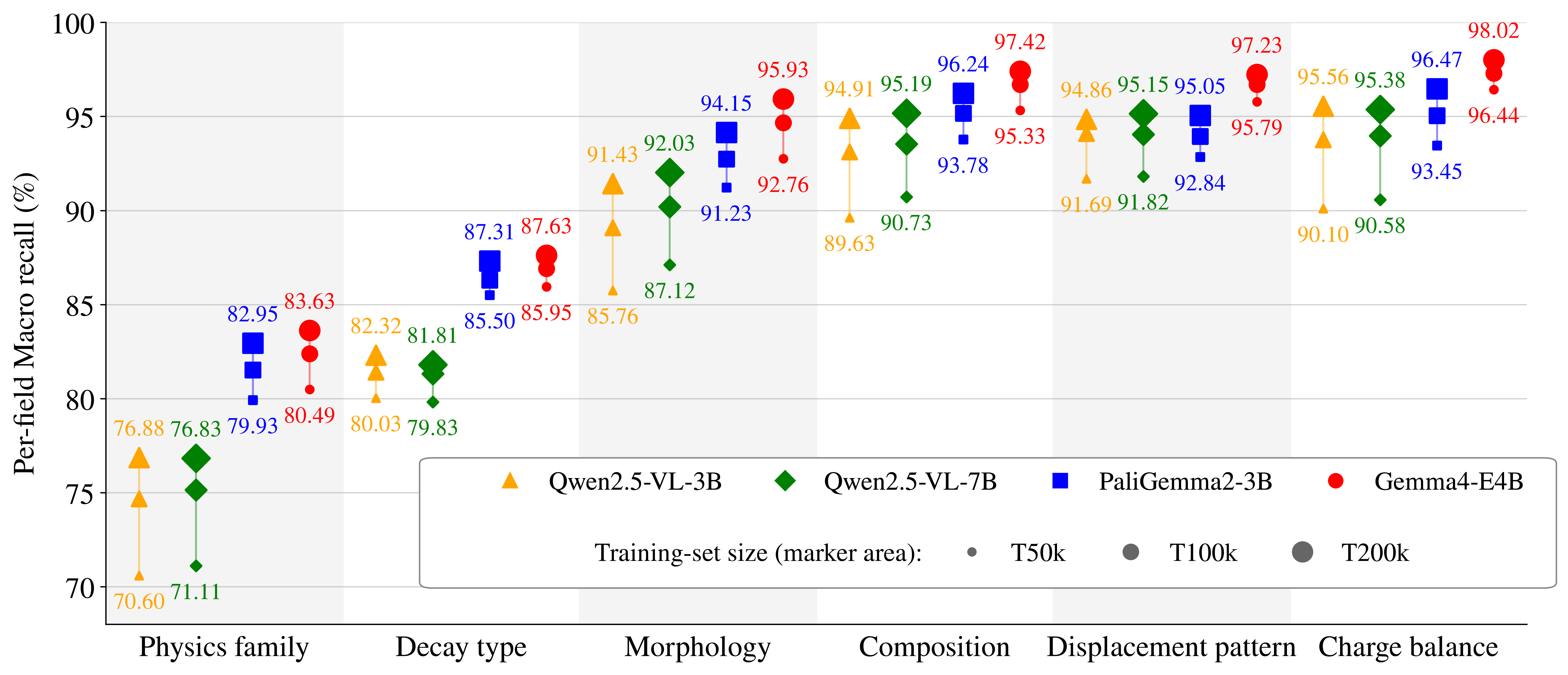}
  \end{minipage}}
  \subfigure[]{
  \begin{minipage}[c]{0.77\linewidth}
  \includegraphics[width=1.0\linewidth]{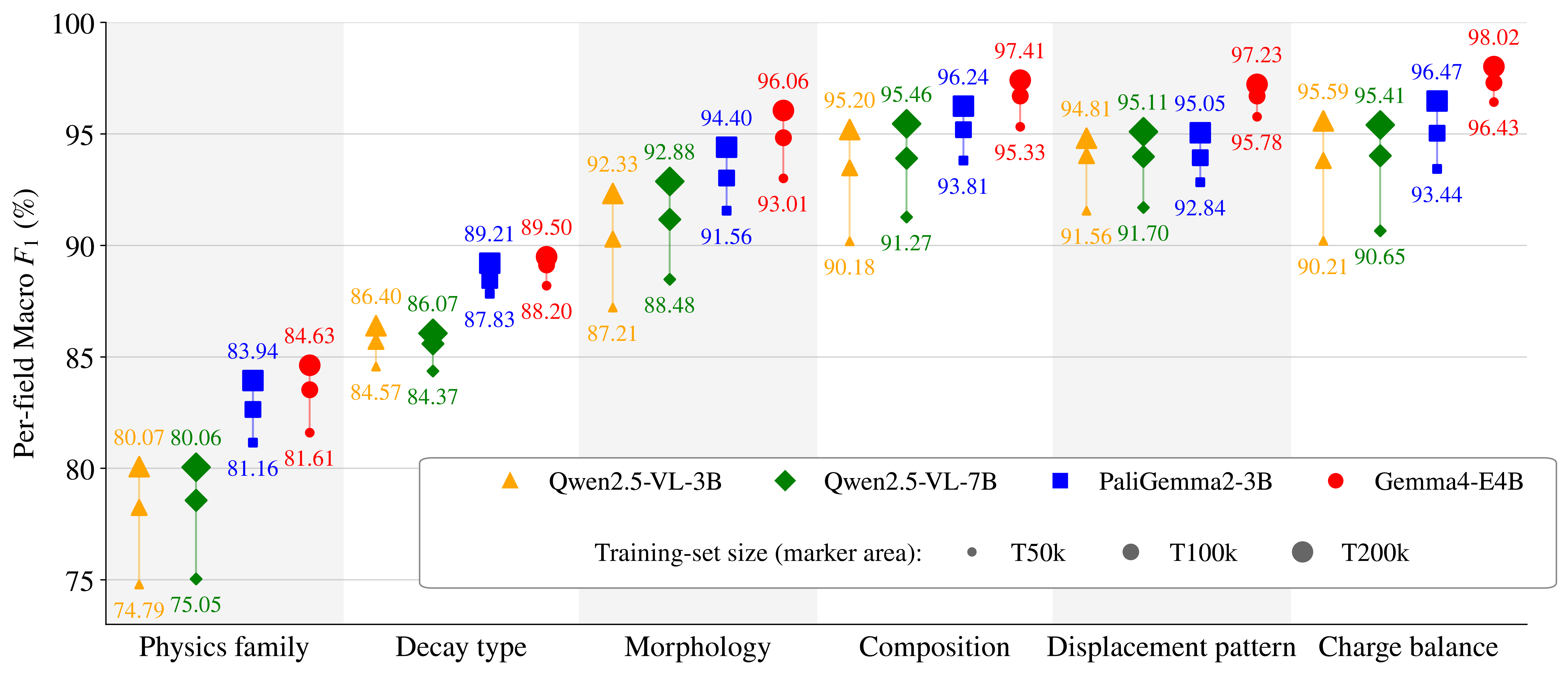}
  \end{minipage}}
  \caption{Task~2 field-wise performance across models and balanced training scales. (a) Per-field Macro recall. (b) Per-field Macro \(F_1\). Colors and marker shapes identify the model, while marker size denotes the training-set size (\Tsmall{}, \Tmedium{}, or \Tlarge{}). All values are percentages.}
  \label{fig:MT_task2}
\end{figure}

\paragraph{Task~2:}
Fig.~\ref{fig:MT_task2} reports (a) per-field Macro recall and (b) per-field Macro \(F_1\). Macro recall gives equal weight to every allowed value, while Macro \(F_1\) summarizes the class-wise precision--recall balance. Both metrics improve monotonically from \Tsmall{} to \Tlarge{} for every model and field. At \Tlarge{}, \GemmaFour{} performs best across all six fields, with \PaliGemma{} generally ranking second. The family and decay-type fields remain the most challenging, whereas the four image-derived attributes achieve higher scores. Numerical values are provided in Appendix~\ref{app:table_task2}.

\begin{figure}[!htpb]\centering
\includegraphics[width=1.0\linewidth]{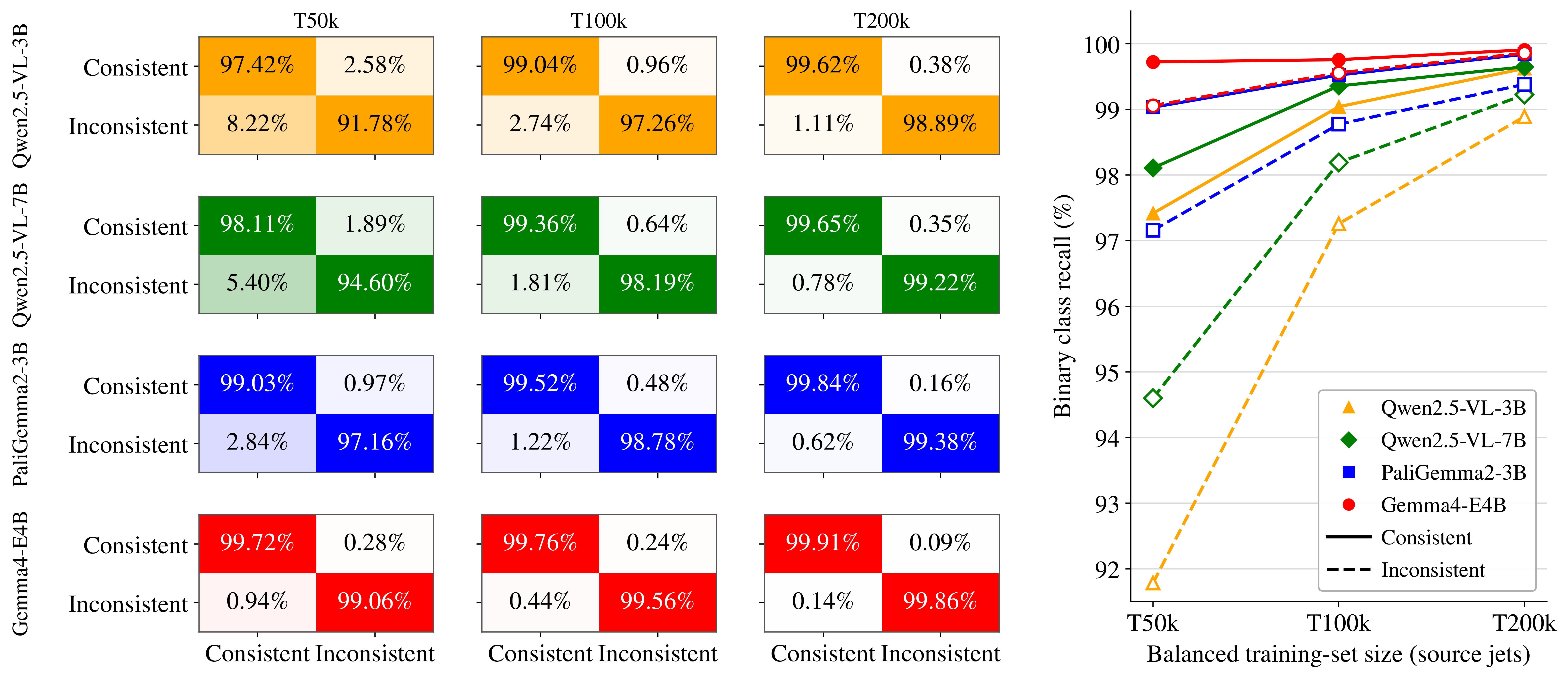}
\caption{Task~3 binary cross-panel consistency performance. The left side shows row-normalized confusion matrices for four models and three balanced training-set sizes, evaluating only whether the four panels are consistent or inconsistent. The right side shows the corresponding class recalls. Colors and markers identify the model; solid and dashed lines denote consistent- and inconsistent-example recall, respectively.}
\label{fig:MT_task3_1}
\end{figure}

\paragraph{Task~3:}
Fig.~\ref{fig:MT_task3_1} evaluates binary cross-panel consistency, distinguishing whether all four panels originate from the same jet or exactly one panel has been replaced. Binary consistency performance improves monotonically with training-set size for all four models. Inconsistent examples are generally more difficult at \Tsmall{}, but the recall gap between the two classes narrows as the training set grows. \GemmaFour{} performs best at every scale and reaches \(99.91\%\) recall for consistent examples and \(99.86\%\) for inconsistent examples at \Tlarge{}, corresponding to \(99.88\%\) binary Macro recall.

Table~\ref{tab:multitask_task3_Paligemma} examines the localization performance of \PaliGemma{} by replacement panel and donor difficulty. Easy replacements are localized with nearly \(100\%\) recall, whereas hard P3 and P4 replacements remain more challenging. From \Tsmall{} to \Tlarge{}, their recall deficits relative to Easy decrease from \(8.60\) to \(2.70\) percentage points for P3 and from \(10.50\) to \(1.80\) percentage points for P4. Thus, difficulty-specific gains remain visible even when aggregate Task~3 performance is near saturation.

\begin{figure}[!htpb]\centering
\includegraphics[width=1.0\linewidth]{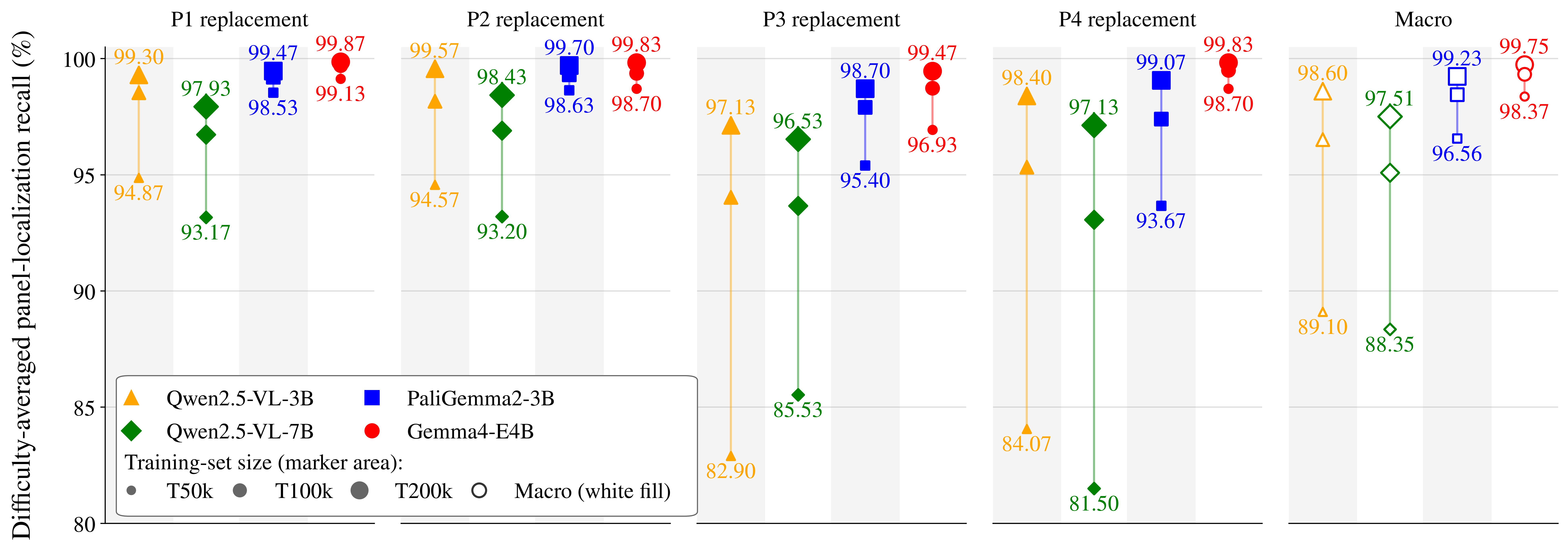}
\caption{Difficulty-averaged panel-localization recall on inconsistent Task~3 examples. Marker area denotes training-set size; white-filled markers show the localization Macro recall over P1--P4.}
\label{fig:MT_task3_2}
\end{figure}

Fig.~\ref{fig:MT_task3_2} extends the difficulty-averaged localization comparison to all four models. The largest scaling gains generally occur for P3 and P4. At \Tlarge{}, localization Macro recall ranges from \(97.51\%\) to \(99.75\%\) across the four models. Detailed data records are in Appendix~\ref{app:table_task3}.
  
\begin{table}[!htpb]\centering\footnotesize
\renewcommand{\arraystretch}{1.0}
\caption{Task~3 localization recall of \PaliGemma{} on the 12,000 inconsistent examples in \test{}. Easy columns give recall, whereas Medium and Hard columns give percentage-point differences relative to Easy. Mean columns average the three donor difficulties, and the final row gives localization Macro recall over P1--P4.}
\label{tab:multitask_task3_Paligemma}
\begin{tabular}{ccccccccccccc}
\hline \hline 
& \multicolumn{4}{c}{T50k} & \multicolumn{4}{c}{T100k}& \multicolumn{4}{c}{T200k}                        \\ \cmidrule(lr){2-5} \cmidrule(lr){6-9} \cmidrule(lr){10-13} 
\multirow{-2}{*}{\begin{tabular}[c]{@{}c@{}}Replaced \\ panel\end{tabular}} 
& Easy  & Medium & Hard & \cellcolor[HTML]{CBCEFB}Mean 
& Easy  & Medium & Hard & \cellcolor[HTML]{CBCEFB}Mean 
& Easy  & Medium & Hard & \cellcolor[HTML]{CBCEFB}Mean \\ \hline
P1 & 99.90 & -1.30 & -2.80 & \cellcolor[HTML]{CBCEFB}98.53 & 100.00 & -0.80 & -1.60 & \cellcolor[HTML]{CBCEFB}99.20 & 100.00 & -0.70 & -0.90 & \cellcolor[HTML]{CBCEFB}99.47 \\
P2 & 99.70 & -0.80 & -2.40 & \cellcolor[HTML]{CBCEFB}98.63 & 99.90 & -0.60 & -1.20 & \cellcolor[HTML]{CBCEFB}99.30 & 100.00 & -0.20 & -0.70 & \cellcolor[HTML]{CBCEFB}99.70 \\
P3 & 99.70 & -4.30 & -8.60 & \cellcolor[HTML]{CBCEFB}95.40 & 100.00 & -2.00 & -4.30 & \cellcolor[HTML]{CBCEFB}97.90 & 100.00 & -1.20 & -2.70 & \cellcolor[HTML]{CBCEFB}98.70 \\
P4 & 99.20 & -6.10 & -10.50 & \cellcolor[HTML]{CBCEFB}93.67 & 99.70 & -1.70 & -5.20 & \cellcolor[HTML]{CBCEFB}97.40 & 99.70 & -0.10 & -1.80 & \cellcolor[HTML]{CBCEFB}99.07 \\
\rowcolor[HTML]{CBCEFB}Macro recall & & & & 96.56 & & & & 98.45 & & & & 99.23 \\ \hline \hline
\end{tabular}
\end{table}

\subsection{Targeted mitigation of $W/Z$ confusion}\label{sec:wz_results}
Tables~\ref{tab:panel_ablation_test24k} and \ref{tab:multitask_task1_test24k} identify \texttt{Zqq} as a persistent Task~1 performance bottleneck, while Fig.~\ref{fig:MT_task1} shows that classification improves as the balanced training set grows. This subsection therefore tests whether \texttt{Zqq} recall can be improved through targeted data reallocation without increasing the total training budget. All configurations contain 200,000 source jets (\Tlarge{}), while the quotas assigned separately to \texttt{Zqq} and \texttt{Wqq} are increased from 20,000 to 30,000 or 40,000, with the remaining class quotas reduced accordingly.

\begin{figure}[!htpb]\centering
\includegraphics[width=1.0\linewidth]{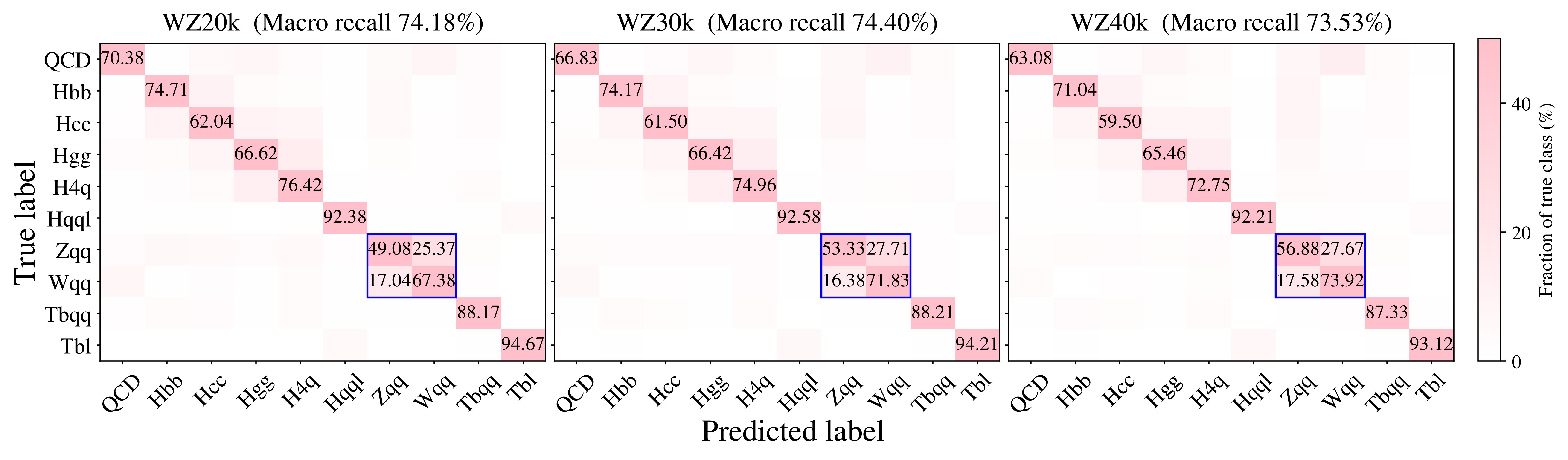}
\caption{Row-normalized Task~1 confusion matrices for \PaliGemma{} under the balanced \Tlarge{} allocation (WZ20k) and the enriched \WZthirty{}/\WZforty{} allocations. Blue boxes highlight the \texttt{Zqq}-\texttt{Wqq} block; titles report Macro recall.}
\label{fig:MT_WZ_task1_CM}
\end{figure}

Fig.~\ref{fig:MT_WZ_task1_CM} compares \PaliGemma{} under the fixed \Tlarge{} training budget. \WZthirty{} provides the highest Macro recall (74.40\%), whereas \WZforty{} further improves \texttt{Zqq} and \texttt{Wqq} recall but reduces performance on several other classes, demonstrating the fixed-budget trade-off.

Table~\ref{tab:wz_task1} extends the representative confusion-matrix comparison to all three models, and \GemmaFour{} is omitted from this comparison because its longer training time made the additional \(W/Z\)-enrichment runs impractical within the available computing budget. The total training budget and \test{} benchmark are fixed. The same trend is observed for all three models. \WZthirty{} provides the highest Macro recall, while \WZforty{} gives the highest \texttt{Zqq} and \texttt{Wqq} recall. Stronger enrichment therefore improves the target categories but eventually reduces aggregate performance because fewer examples remain for the other eight classes.

\begin{table}[!htpb]\centering\footnotesize
\renewcommand{\arraystretch}{1.0}
\caption{Effect of fixed-budget \(W/Z\) class reallocation on Task~1 performance. The total training set contains 200,000 source jets in every configuration. The \(W/Z\) quota denotes the number of training jets assigned separately to each of the \texttt{Zqq} and \texttt{Wqq} categories.}
\label{tab:wz_task1}
\begin{tabular}{clrrr}
\hline \hline
Model & Allocation & Macro recall (\%) & \texttt{Zqq} recall (\%) & \texttt{Wqq} recall (\%)  \\
\hline
\multirow{3}{*}{\QwenThree{}} 
& WZ20k (balanced case) & 72.21 & 44.17 & 65.46 \\
& WZ30k & \textbf{72.32} & 49.96 & 70.71 \\
& WZ40k & 71.97 & \textbf{54.00} & \textbf{72.04} \\
\hline
\multirow{3}{*}{\QwenSeven{}} 
& WZ20k (balanced case) & 73.13 & 46.96 & 65.38 \\
& WZ30k & \textbf{73.22} & 50.58 & 71.29 \\
& WZ40k & 72.55 & \textbf{56.63} & \textbf{72.21} \\
\hline
\multirow{3}{*}{\PaliGemma{}} 
& WZ20k (balanced case) & 74.18 & 49.08 & 67.38 \\
& WZ30k & \textbf{74.40} & 53.33 & 71.83 \\
& WZ40k & 73.53 & \textbf{56.88} & \textbf{73.92} \\
\hline \hline
\end{tabular}
\end{table}

Table~\ref{tab:wz_task23} further compares whether \(W/Z\) enrichment affects Tasks~2 and 3. Task~2 is evaluated using the field-mean Macro recall and field-mean Macro \(F_1\) across its six fields, while Task~3 is evaluated using binary Macro recall and localization Macro recall. Values in parentheses denote changes in percentage points relative to the balanced configuration. (i) Increasing the \(W/Z\) allocation produces a consistent reduction in Task~2 performance. At \WZforty{}, field-mean Macro recall decreases by \(2.41-2.94\) percentage points and field-mean Macro \(F_1\) by \(2.29-3.11\) percentage points across the three models. (ii) In contrast, Task~3 performance remains near saturation: binary Macro recall changes by at most \(0.50\) percentage points, while localization Macro recall changes range from \(-0.15\) to \(+1.09\) percentage points. Thus, \(W/Z\) enrichment does not improve the auxiliary attribute task and has no systematic effect on cross-panel consistency learning.

\begin{table}[!htpb]\centering\footnotesize
\renewcommand{\arraystretch}{1.0}
\caption{Task~2 and Task~3 performance under fixed-budget \(W/Z\) enrichment. Task~2 metrics are field means over the six attributes, while Task~3 localization Macro recall is evaluated on the 12,000 inconsistent examples. All metrics are percentages; parentheses give percentage-point differences from WZ20k. Blue and red shading denote increases and decreases, respectively.}
\label{tab:wz_task23}
\begin{tabular}{cccc cc}
\hline \hline
\multirow{2}{*}{Model} & \multirow{2}{*}{Allocation} & \multicolumn{2}{c}{Task~2} & \multicolumn{2}{c}{Task~3} \\
\cmidrule(lr){3-4}\cmidrule(lr){5-6}
 & & Macro recall & Macro \(F_1\) & Binary Macro recall & Localization Macro recall \\
\hline
\multirow{3}{*}{\QwenThree{}} & WZ20k  & 89.33 & 90.73 & 99.26 & 98.60 \\
 & WZ30k  & \cellcolor[HTML]{FFE2E0}88.46~$(-0.87)$ & \cellcolor[HTML]{FFE2E0}89.79~$(-0.94)$ & \cellcolor[HTML]{D8ECFF}99.38~$(+0.12)$ & \cellcolor[HTML]{D8ECFF}98.83~$(+0.23)$ \\
 & WZ40k & \cellcolor[HTML]{FFE2E0}86.39~$(-2.94)$ & \cellcolor[HTML]{FFE2E0}87.62~$(-3.11)$ & \cellcolor[HTML]{D8ECFF}99.41~$(+0.15)$ & \cellcolor[HTML]{FFE2E0}98.45~$(-0.15)$ \\
\hline
\multirow{3}{*}{\QwenSeven{}} & WZ20k  & 89.40 & 90.83 & 99.44 & 97.51 \\
 & WZ30k & \cellcolor[HTML]{FFE2E0}88.48~$(-0.92)$ & \cellcolor[HTML]{FFE2E0}89.81~$(-1.03)$ & \cellcolor[HTML]{FFE2E0}98.94~$(-0.50)$ & \cellcolor[HTML]{D8ECFF}98.05~$(+0.54)$ \\
 & WZ40k & \cellcolor[HTML]{FFE2E0}86.63~$(-2.77)$ & \cellcolor[HTML]{FFE2E0}87.85~$(-2.98)$ & \cellcolor[HTML]{FFE2E0}99.25~$(-0.19)$ & \cellcolor[HTML]{D8ECFF}98.60~$(+1.09)$ \\
\hline
\multirow{3}{*}{\PaliGemma{}}  & WZ20k  & 92.03 & 92.55 & 99.61 & 99.23 \\
 & WZ30k & \cellcolor[HTML]{FFE2E0}91.23~$(-0.80)$ & \cellcolor[HTML]{FFE2E0}91.83~$(-0.72)$ & \cellcolor[HTML]{FFE2E0}99.59~$(-0.02)$ & \cellcolor[HTML]{D8ECFF}99.30~$(+0.07)$ \\
 & WZ40k & \cellcolor[HTML]{FFE2E0}89.62~$(-2.41)$ & \cellcolor[HTML]{FFE2E0}90.27~$(-2.29)$ & \cellcolor[HTML]{D8ECFF}99.65~$(+0.04)$ & \cellcolor[HTML]{D8ECFF}99.30~$(+0.07)$ \\
\hline \hline
\end{tabular}

\end{table}

\subsection{Transfer to JetClass-II and Aspen Open Jets}\label{sec:transfer_two}
JetClass-II extends the original JetClass benchmark to a broader collection of large-radius jet signatures~\cite{li2024accelerating, jetclass2}. Its samples are grouped into three coarse categories: \texttt{QCD} for multijet background, \texttt{Res2P} for generic two-prong resonant jets, and \texttt{Res34P} for generic three- or four-prong resonant jets. The Aspen Open Jets dataset contains approximately 178 million high $\pt$ jets reconstructed from the 2016 CMS JetHT Open Data~\cite{amram2025aspen}. In contrast to the simulated JetClass datasets, it provides a real-data stress test with detector and reconstruction effects. For each target dataset, disjoint subsets of 3,000, 600, and 9,000 jets are used for training, validation, and testing, respectively.

Task~1 is not evaluated on Aspen Open Jets because collision data do not provide reliable per-jet truth-category labels. In particular, the available ParticleNet scores are not treated as ground truth, since doing so would replace an independent transfer test with labels inherited from another trained classifier. Therefore, the Aspen evaluation retains only Task~2 and Task~3.

Transfer is evaluated using the \PaliGemma{} and \GemmaFour{} \Tlarge{} multitask adapters, which are selected as the two strongest \Tlarge{} configurations across the source-domain tasks. Three settings are considered: an analytic chance baseline, direct transfer of the frozen JetClass adapter, and limited target-domain adaptation. The chance baseline is the expected performance under uniform sampling over the valid task-specific output states and does not involve model inference. Direct transfer evaluates the JetClass adapter without target-domain training, whereas target-adapted transfer continues task-specific LoRA training on Train3k.

Table~\ref{tab:jetclass2_transfer_v1} shows that target adaptation substantially improves Task~1. Macro recall increases from \(34.17\%\) to \(70.80\%\) for \PaliGemma{} and from \(56.38\%\) to  \(70.72\%\) for \GemmaFour{}. Direct Task~3 transfer is already strong for both models. Task~2 transfer is model-dependent: Direct \GemmaFour{} reaches \(96.13\%\) field-mean Macro recall, whereas \PaliGemma{} increases from \(8.21\%\) under Direct transfer to \(95.08\%\) after target adaptation. These results indicate that the new Task~1 label space benefits from limited adaptation, while Task~3 transfers effectively without target-domain training.

Table~\ref{tab:aspen_transfer_v1} evaluates transfer to real collision data. Direct \GemmaFour{} transfer reaches \(95.14\%\) field-mean Macro recall and \(95.36\%\) field-mean Macro \(F_1\) on Task~2, with only modest changes after target adaptation. For \PaliGemma{}, target adaptation raises these metrics from \(24.95\%\) and \(38.92\%\) to \(92.89\%\) and \(93.42\%\), respectively. Task~3 transfers effectively under the Direct condition. Thus, the image-derived tasks can transfer from simulated JetClass jets to real-data images, although the benefit of limited target adaptation depends on the model and task.
  
\begin{table}[!htpb]\centering\footnotesize
\renewcommand{\arraystretch}{1.0}
\caption{Transfer of the \Tlarge{} JetClass adapters to JetClass-II. Chance is the analytic expectation under uniform sampling over valid task outputs; Direct evaluates the frozen source adapter; Target-adapted continues task-specific training on Train3k and selects the checkpoint on Val600. Direct and Target-adapted are evaluated on the same Test9k set, with Task~3 localization evaluated on its 4,500 inconsistent examples. Parentheses give Target-adapted minus Direct differences in percentage points; all entries are percentages.}
\label{tab:jetclass2_transfer_v1}
\begin{tabular}{ccccccc}
\hline \hline
& & & \multicolumn{2}{c}{\PaliGemma{}} & \multicolumn{2}{c}{\GemmaFour{}} \\ \cline{4-7} 
\multirow{-2}{*}{Task}   
& \multirow{-2}{*}{Metric}                 
& Chance  & Direct  & Target-adapted  & Direct  & Target-adapted 
\\ \hline
& \texttt{QCD} recall & 33.33 & 99.90 & 61.20 $(-38.70)$ & 81.23 & 60.80 $(-20.43)$ \\
& \texttt{Res2P} recall & 33.33 & 0.23 & 70.70 $(+70.47)$ & 51.03 & 76.93 $(+25.90)$ \\
& \texttt{Res34P} recall & 33.33 & 2.37 & 80.50 $(+78.13)$ & 36.87 & 74.43 $(+37.57)$ \\
& \cellcolor[HTML]{CBCEFB}Macro recall & \cellcolor[HTML]{CBCEFB}33.33 & \cellcolor[HTML]{CBCEFB}34.17 & \cellcolor[HTML]{CBCEFB}70.80 $(+36.63)$ & \cellcolor[HTML]{CBCEFB}56.38 & \cellcolor[HTML]{CBCEFB}70.72 $(+14.34)$ \\
& \texttt{QCD} \(F_1\) & 33.33 & 50.40 & 70.89 $(+20.48)$ & 62.59 & 71.31 $(+8.71)$ \\
& \texttt{Res2P} \(F_1\) & 33.33 & 0.47 & 68.28 $(+67.81)$ & 54.94 & 68.63 $(+13.69)$ \\
& \texttt{Res34P} \(F_1\) & 33.33 & 4.58 & 73.10 $(+68.52)$ & 47.67 & 72.52 $(+24.85)$ \\
\multirow{-8}{*}{Task 1} 
& \cellcolor[HTML]{CBCEFB}Macro \(F_1\) & \cellcolor[HTML]{CBCEFB}33.33 & \cellcolor[HTML]{CBCEFB}18.48 & \cellcolor[HTML]{CBCEFB}70.76 $(+52.27)$ & \cellcolor[HTML]{CBCEFB}55.07 & \cellcolor[HTML]{CBCEFB}70.82 $(+15.75)$ \\ \hline
& Field-mean Macro recall & 29.17 & 8.21 & 95.08 $(+86.87)$ & 96.13 & 95.93 $(-0.20)$ \\
\multirow{-2}{*}{Task 2} 
& Field-mean Macro \(F_1\) & 28.58 & 15.04 & 95.13 $(+80.09)$ & 96.11 & 96.14 $(+0.03)$ \\ \hline
& \cellcolor[HTML]{CBCEFB}Binary Macro recall  & \cellcolor[HTML]{CBCEFB}50.00 & \cellcolor[HTML]{CBCEFB}98.73 & \cellcolor[HTML]{CBCEFB}98.79 $(+0.06)$  & \cellcolor[HTML]{CBCEFB}99.34 & \cellcolor[HTML]{CBCEFB}99.42 $(+0.08)$  \\
& Consistent recall & 20.00 & 99.53 & 99.60 $(+0.07)$ & 99.89 & 99.91 $(+0.02)$ \\
& Inconsistent recall & 80.00 & 97.93 & 97.98 $(+0.04)$ & 98.80 & 98.93 $(+0.13)$ \\
\multirow{-4}{*}{Task 3} 
& \cellcolor[HTML]{CBCEFB}Localization Macro recall & \cellcolor[HTML]{CBCEFB}20.00 & \cellcolor[HTML]{CBCEFB}97.36 & \cellcolor[HTML]{CBCEFB}97.62 $(+0.27)$ & \cellcolor[HTML]{CBCEFB}98.11 & \cellcolor[HTML]{CBCEFB}98.53 $(+0.42)$ \\ 
\hline \hline
\end{tabular}
\end{table}

\begin{table}[!htpb]\centering\footnotesize
\renewcommand{\arraystretch}{1.0}
\caption{Transfer of the \Tlarge{} JetClass adapters to Aspen Open Jets. The transfer settings, evaluation protocol, and parenthetical differences follow Table~\ref{tab:jetclass2_transfer_v1}; all entries are percentages.}
\label{tab:aspen_transfer_v1}
\begin{tabular}{ccccccc}
\hline \hline
& & & \multicolumn{2}{c}{\PaliGemma{}} & \multicolumn{2}{c}{\GemmaFour{}} \\ \cline{4-7} 
\multirow{-2}{*}{Task}   
& \multirow{-2}{*}{Metric}                      
& Chance  & Direct  & Target-adapted  & Direct  & Target-adapted                               \\ \hline
& Field-mean Macro recall 
& 29.17  & 24.95  & 92.89 $(+67.93)$  & 95.14  & 95.64 $(+0.50)$                         \\
\multirow{-2}{*}{Task 2} 
& Field-mean Macro \(F_1\)                           
& 27.72  & 38.92  & 93.42 $(+54.50)$  & 95.36  & 95.92 $(+0.56)$                         \\ \hline
& \cellcolor[HTML]{CBCEFB}Binary Macro recall       
& \cellcolor[HTML]{CBCEFB}50.00 & \cellcolor[HTML]{CBCEFB}97.48 & \cellcolor[HTML]{CBCEFB}98.88 $(+1.40)$ & \cellcolor[HTML]{CBCEFB}99.87 & \cellcolor[HTML]{CBCEFB}99.93 $(+0.07)$ \\
& Consistent recall                           
& 20.00  & 96.67  & 99.44 $(+2.78)$  & 99.78  & 99.98 $(+0.20)$                         \\
& Inconsistent recall                         
& 80.00  & 98.29  & 98.31 $(+0.02)$  & 99.96  & 99.89 $(-0.07)$                         \\
\multirow{-4}{*}{Task 3} 
& \cellcolor[HTML]{CBCEFB}Localization Macro recall & \cellcolor[HTML]{CBCEFB}20.00 & \cellcolor[HTML]{CBCEFB}97.29 & \cellcolor[HTML]{CBCEFB}97.64 $(+0.36)$ & \cellcolor[HTML]{CBCEFB}99.82 & \cellcolor[HTML]{CBCEFB}99.82 $(+0.00)$ \\ \hline \hline
\end{tabular}
\end{table}

\section{Conclusion}\label{sec:conclusion}
This work establishes several analyses and quantitative findings for instruction-driven jet analysis with vision-language models. Generic VLM pretraining alone is insufficient: prompted zero-shot Macro recall remains at or below the ten-class chance level (\(3.33\%\)--\(10.57\%\)). Parameter-efficient adaptation removes this deficit, and Task~1 Macro recall rises monotonically with training-set size for all four models, reaching \(75.05\%\) (Macro \(F_1\) \(74.87\%\)) for \GemmaFour{} at T200k with only \(0.2\%\) of the JetClass training partition. Panel ablations quantify where the discrimination originates.
Relative to transverse-momentum composition alone, particle-species and track-displacement information drive the largest gains, raising \PaliGemma{} Task~1 Macro recall from \(57.50\%\) to \(68.80\%\). The displacement panel supplies the heavy-flavor lifetime signature (i.e. the displaced secondary vertices and nonzero impact parameters left by the finite lifetimes of b and c hadrons), while the charge panel contributes the complementary information that separates the hadronic \(W\) and \(Z\) whose \(\sim\!11\)~GeV mass splitting is comparable to the jet-mass resolution and therefore leaves the two resonance peaks overlapping. Multitask adaptation adds further \(1.41\) points to \(70.21\%\). Beyond classification, the same adapted model reaches \(93.31\%\) Task~2 field-mean Macro recall, with binary and localization Macro recall reaching \(99.88\%\)/\(99.75\%\) correspondingly. Physically, this means the considered model not only tags the jet but also recovers the underlying event content—the number of prongs, the presence of leptons, and the heavy-flavor composition that define each decay topology, demonstrating that the learned representation captures the physics of the jet rather than a surface-level classification shortcut. A fixed-budget reallocation toward \(W/Z\) jets improves the difficult \texttt{Zqq} recall by up to \(9.83\) points at nearly constant Macro recall, quantifying an error-guided data trade-off. Finally, the four-panel representation transfers across domains: with only 3{,}000 target-domain jets, lightweight adaptation exceeds \(70\%\) Task~1 Macro recall on JetClass-II and surpasses \(92\%\) on every reported Task~2 and Task~3 metric on both JetClass-II and the real-data Aspen Open Jets, with Task~3 and \GemmaFour{} Task~2 already strong under frozen transfer.

\section*{Acknowledgments}

The authors would like to thank the Research Computing Department at Khalifa University for the computational resources used in this study. This work was carried out using the high-performance computing facilities provided by the university, and the authors gratefully acknowledge access to the facility's NVIDIA H200 GPU resources, which provided the crucial computing support for this study.
LZ and RS gratefully acknowledge the support of this research work by the FSU grant "Search for New Physics Beyond the Standard Model at Collider Physics" (N. 8474000768) and the International Research Network Expansion (IRNE) grant (N. 8471000010). RS would like to thank CERN for the hospitality while part of this work was carried out.

\bibliographystyle{unsrt}
\bibliography{references}

\appendix
\begin{appendices}
\section{User Instructions and Output Schemas}\label{app:prompts}
This appendix reproduces the task-specific user messages used in the reported experiments. For each model, the corresponding processor inserts the image placeholder and applies the model-specific chat template; these wrapper elements are omitted here. Box titles and visual formatting, including colors, shadows, and automatic line wrapping, are typographic only and are not supplied to the models. The short Task~1 instruction is used both during adaptation and for the strict zero-shot control. The label-and-schema prompted zero-shot control uses the separate explicit instruction reproduced below.

The term ``energy'' is retained where it appeared in the original Task~2 and Task~3 instructions. It refers to P1, which encodes transverse-momentum composition as defined in Section~\ref{sec:four_panel_representation}.

\subsection{Instructions for adapted models}
\begin{instructionbox}{Task~1 instruction}
What is the JetClass truth label?
\end{instructionbox}

\begin{instructionbox}{Task~2 instruction}
Analyze the qualitative properties visible in this 2x2 four-panel JetClass image. The panels are energy composition (top left), particle species (top right), displacement (bottom left), and charge structure (bottom right). Return only valid compact JSON with exactly these fields and allowed values: 
family=[qcd,higgs,vector_boson,top]; 
decay_type=[not_applicable,hadronic,semileptonic]; 
morphology=[single_core,double_core,multicore,diffuse]; 
composition=[charged_dominated,neutral_dominated,em_enriched,mixed]; 
displacement_pattern=[weak,localized,diffuse]; 
charge_balance=[positive_skewed,balanced,negative_skewed].
\end{instructionbox}

\begin{instructionbox}{Task~3 instruction}
Determine whether the energy, species, displacement, and charge panels in this 2x2 JetClass image all come from the same jet. If one panel is inconsistent, identify it. Return only valid compact JSON with all_panels_consistent=true or false and mismatched_panel equal to one of [none,energy,species,displacement,charge].
\end{instructionbox}

\subsection{Label-and-schema prompted zero-shot Task~1}
For the label-and-schema prompted zero-shot control, the short Task~1 instruction is replaced by the following explicit instruction.
\begin{instructionbox}{Task~1 zero-shot instruction}
Classify this 2x2 four-panel JetClass image. Choose exactly one truth label from [QCD,Hbb,Hcc,Hgg,H4q,Hqql,Zqq,Wqq,Tbqq,Tbl]. Return only valid compact JSON with exactly one field using this schema: {"label":"<chosen_label>"}. The value of label must be one of the allowed labels. Do not include an explanation, Markdown fencing, or any additional fields.
\end{instructionbox}

\section{Deterministic Construction of Task~2 Targets}\label{app:task2_targets}
Task~2 contains two truth-label-derived fields and four rendered-image-derived fields. All image descriptors are computed from the normalized \(224\times224\) RGB raster supplied to the model processor. For each processed dataset, quantile thresholds are fitted on its training images and then frozen for testing.

\subsection{Truth-label-derived attributes}
The deterministic mapping from the Task~1 category to \texttt{family} and \texttt{decay\_type} is listed in Table~\ref{tab:truth_hierarchy}. These fields are target definitions, not attributes inferred independently from image morphology.

\begin{table}[!htpb]\centering\footnotesize
\caption{Deterministic mapping from Task~1 truth categories to the two truth-label-derived Task~2 fields.}
\label{tab:truth_hierarchy}
\begin{tabular}{lll}
\hline \hline
Task-1 label(s) & \texttt{family} & \texttt{decay\_type} \\
\hline
\texttt{QCD}                                               & qcd            & not\_applicable \\
\texttt{Hbb}, \texttt{Hcc}, \texttt{Hgg}, \texttt{H4q}     & higgs          & hadronic \\
\texttt{Hqql}                                              & higgs          & semileptonic \\
\texttt{Zqq}, \texttt{Wqq}                                 & vector\_boson  & hadronic \\
\texttt{Tbqq}                                              & top            & hadronic \\
\texttt{Tbl}                                               & top            & semileptonic \\
\hline \hline
\end{tabular}
\end{table}

\subsection{Rendered-image-derived attributes}
Let \(I_{p,c}(u,v)\) be the stored 8-bit intensity defined in Section~\ref{sec:four_panel_representation}. The rescaled intensity and its channel integral are
\begin{equation}
\widetilde I_{p,c}(u,v)=\frac{I_{p,c}(u,v)}{255},
\qquad
\mathcal I_{p,c}=\sum_{u,v}\widetilde I_{p,c}(u,v)
\end{equation}
where \(p\) and \(c\) denote the panel and RGB channel, respectively. A numerical floor \(\epsilon_{\mathrm{attr}}=10^{-8}\) is used in the descriptor denominators below.

\paragraph{Morphology.}
For core counting only, \(\widetilde I_{\mathrm{P1},R}\) is smoothed by two \(3\times3\) mean-filter passes. Local maxima at least 30\% of the smoothed global peak are retained using a nine-pixel non-maximum-suppression radius, with at most four retained cores. The remaining morphology descriptors are computed from \(\widetilde I_{\mathrm{P1},R}\) without this additional mean filtering. Let \(n_{\mathrm{pix}}=112^2\), let \(f_{\mathrm{act}}\) be the fraction of its pixels above 0.05, and let \(f_{\mathrm{top10}}\) be the fraction of the total \(\widetilde I_{\mathrm{P1},R}\) intensity contained in its brightest \(\lceil0.1n_{\mathrm{pix}}\rceil\) pixels. The diffuse score is
\begin{equation}
S_{\mathrm{diff}}=\frac{f_{\mathrm{act}}}{\max(f_{\mathrm{top10}},\epsilon_{\mathrm{attr}})}
\end{equation}
Values above the training-set 75th percentile are \texttt{diffuse}. Otherwise, zero or one core gives \texttt{single\_core}, two gives \texttt{double\_core}, and at least three gives \texttt{multicore}.

\paragraph{Composition.}
The rendered charged-channel and P2-B fractions are
\begin{equation}
f_{\mathrm{ch}}
=\frac{\mathcal I_{\mathrm{P1},G}}
{\max\!\left(\mathcal I_{\mathrm{P1},G}+\mathcal I_{\mathrm{P1},B},\epsilon_{\mathrm{attr}}\right)},
\qquad
f_{\mathrm{em}}
=\frac{\mathcal I_{\mathrm{P2},B}}
{\max\!\left(\mathcal I_{\mathrm{P2},R}+\mathcal I_{\mathrm{P2},G}+\mathcal I_{\mathrm{P2},B},\epsilon_{\mathrm{attr}}\right)}
\end{equation}
If $f_{\mathrm{em}}$ exceeds its training-set 80th percentile, the target is \texttt{em\_enriched}. Otherwise, $f_{\mathrm{ch}}$ above the two-thirds quantile gives \texttt{charged\_dominated}, below the one-third quantile gives \texttt{neutral\_dominated}, and the intermediate range gives \texttt{mixed}.

\paragraph{Displacement pattern.}
At each pixel, the maximum of the three rescaled P3 intensities \(\widetilde I_{\mathrm{P3},c}\) is taken, and \(f_{\mathrm{disp}}\) is the fraction of these values above 0.05. Values at or below the training-set one-third quantile are \texttt{weak}; values above the two-thirds quantile are \texttt{diffuse}; intermediate values are \texttt{localized}.

\paragraph{Charge balance.}
The signed rendered charge skew is defined as
\begin{equation}
S_{\mathrm{ch}}
=\frac{\mathcal I_{\mathrm{P4},R}-\mathcal I_{\mathrm{P4},G}}
{\max\!\left(\mathcal I_{\mathrm{P4},R}+\mathcal I_{\mathrm{P4},G},\epsilon_{\mathrm{attr}}\right)}
\end{equation}
Values above the training-set two-thirds quantile are \texttt{positive\_skewed}, values below the one-third quantile are \texttt{negative\_skewed}, and intermediate values are \texttt{balanced}.

\section{Construction of Task~3 Panel-Replacement Examples}\label{app:task3_construction}
Task~3 is constructed independently within the training, validation, and test splits. Within each truth class, half of the source jets are assigned to the consistent subset and retain their original images, with targets \texttt{all\_panels\_consistent=true} and \texttt{mismatched\_panel=none}. The remaining jets are assigned to the inconsistent subset, in which exactly one panel is copied from a distinct donor jet in the same official split. Their targets are \texttt{all\_panels\_consistent=false} and the identifier of the replaced panel.

\paragraph{Difficulty strata.}
Donor difficulty controls construction but is not a prediction target. Jet-$\pt$ and soft-drop-mass quartile bins used for matching are fitted on the corresponding training set and then applied to test.

\begin{itemize}
  \item \textbf{Easy:} the donor has a different truth family and, when possible, a different morphology; selection favors the largest difference in the replaced panel.
  \item \textbf{Medium:} the donor has a different truth label. Candidate pools progressively relax matches in family, morphology, jet-$\pt$ bin, and mass bin before falling back to any different-label jet.
  \item \textbf{Hard:} the donor has the same truth label. Candidate pools progressively relax matching from label, morphology, jet-$\pt$ bin, and mass bin to the label alone.
\end{itemize}

\paragraph{Donor ranking.}
Within each difficulty stratum, at most 24 eligible donor candidates are evaluated in a deterministic order generated with seed 42. Candidate panels are compared using compact image descriptors that summarize intensity, active area, channel composition, spatial centroid, and radial spread. Donor selection favors a visibly different replacement panel while keeping the other three panels as similar as possible. For the medium and hard strata, candidates satisfying a minimum replacement-panel difference are preferred when available. Remaining ties are resolved deterministically by candidate index.

\section{End-to-End Multitask Inference Example}\label{app:demo}
This appendix presents an end-to-end inference example illustrating the common image--instruction interface. The accompanying inference utility accepts a four-panel image and any compatible model-specific LoRA adapter, and returns schema-validated \JSON{} predictions for the three tasks. 

For illustration, a synthetic inconsistent example constructed from jets in the official JetClass test partition is evaluated using the \GemmaFour{} adapter trained on \Tlarge{}. Its displacement panel P3 is replaced by the corresponding panel of a donor jet. The same composite image is paired with all three task instructions to illustrate the shared inference interface.

\begin{examplebox}{End-to-end example}
\begin{center}
\text{Visual input}\\[0.3em]
\includegraphics[width=0.25\linewidth]{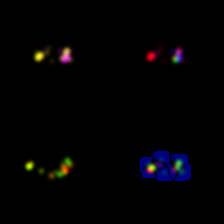}
\end{center}

\begin{tabularx}{\linewidth}{
  @{}
  >{\bfseries\raggedright\arraybackslash}p{2.5cm}
  >{\raggedright\arraybackslash}X
  @{}
}
Adapter    : & \path{gemma4_E4B_T200k_V5k_224px_4panel_VLM_train_1000073}  \\[0.3em]
Base model : & \path{google/gemma-4-E4B-it}                                \\[0.3em]
Backend    : & \texttt{gemma4}                                             \\[0.3em]
Image      : & \path{data/demo/test_001602.png}        
\end{tabularx}

\medskip
\hrule
\medskip

\noindent\textbf{Task~1: truth-class prediction}

\begin{tabularx}{\linewidth}{
  @{}
  >{\bfseries\raggedright\arraybackslash}p{2.8cm}
  >{\raggedright\arraybackslash}X
  @{}
}
Instruction :
&
{\ttfamily\small
What is the JetClass truth label?
}
\\[0.6em]

Output :
&
\texttt{\{"label":"Hgg"\}}
\\[0.3em]

Schema valid :
&
\text{Yes}
\end{tabularx}

\medskip
\hrule
\medskip

\noindent\textbf{Task~2: structured attributes}

\begin{tabularx}{\linewidth}{
  @{}
  >{\bfseries\raggedright\arraybackslash}p{2.8cm}
  >{\raggedright\arraybackslash}X
  @{}
}
Instruction :
&
{\ttfamily\small
Analyze the qualitative properties visible in this 2x2 four-panel JetClass image. The panels are energy composition (top left), particle species (top right), displacement (bottom left), and charge structure (bottom right). Return only valid compact JSON with exactly these fields and allowed values:
family=[qcd,higgs,vector\_boson,top];
decay\_type=[not\_applicable,hadronic,semileptonic];
morphology=[single\_core,double\_core,multicore,diffuse];
composition=[charged\_dominated,neutral\_dominated,em\_enriched,mixed];
displacement\_pattern=[weak,localized,diffuse];
charge\_balance=[positive\_skewed,balanced,negative\_skewed].
}
\\[0.8em]

Output :
&
\texttt{%
\{"family":"higgs",\allowbreak%
"decay\_type":"hadronic",\allowbreak%
"morphology":"multicore",\allowbreak%
"composition":"mixed",\allowbreak%
"displacement\_pattern":"diffuse",\allowbreak%
"charge\_balance":"negative\_skewed"\}%
}
\\[0.3em]

Schema valid :
&
\text{Yes}
\end{tabularx}

\medskip
\hrule
\medskip

\noindent\textbf{Task~3: panel consistency}

\begin{tabularx}{\linewidth}{
  @{}
  >{\bfseries\raggedright\arraybackslash}p{2.8cm}
  >{\raggedright\arraybackslash}X
  @{}
}
Instruction :
&
{\ttfamily\small
Determine whether the energy, species, displacement, and charge panels in this 2x2 JetClass image all come from the same jet. If one panel is inconsistent, identify it. Return only valid compact JSON with all\_panels\_consistent=true or false and mismatched\_panel equal to one of [none,energy,species,displacement,charge].
}
\\[0.8em]

Output :
&
\texttt{%
\{"all\_panels\_consistent":false,\allowbreak%
"mismatched\_panel":"displacement"\}%
}
\\[0.3em]

Schema valid :
&
\text{Yes}
\end{tabularx}

\end{examplebox}

\section{Additional tables for Section~\ref{sec:scalinglaw_balanced} (Task~2)}\label{app:table_task2}
Tables~\ref{tab:appendix_table_task2_1} and \ref{tab:appendix_table_task2_2} provide the numerical Task~2 results underlying Fig.~\ref{fig:MT_task2}.

\begin{table}[!htpb]\centering\footnotesize
\renewcommand{\arraystretch}{1.5}
\caption{Task~2 per-field Macro recall (\%) on \test{}. Per-field values are unweighted over their allowed categories; Field mean is the unweighted average over the six fields.}
\label{tab:appendix_table_task2_1}
\begin{tabular}{cc rr rrrr c}
\hline \hline
\multirow{2}{*}{Model} & \multirow{2}{*}{Training set} & \multicolumn{2}{c}{Truth-label-derived} & \multicolumn{4}{c}{Rendered-image-derived} & \multirow{2}{*}{Field mean} \\
\cmidrule(lr){3-4} \cmidrule(lr){5-8}
 & & Family & Decay & Morph. & Comp. & Displ. & Charge & \\
\hline
\multirow{3}{*}{\QwenThree{}} 
     & T50k & 70.60 & 80.03 & 85.76 & 89.63 & 91.69 & 90.10 & \cellcolor[HTML]{CBCEFB}84.64 \\
     & T100k & 74.68 & 81.41 & 89.09 & 93.12 & 94.09 & 93.76 & \cellcolor[HTML]{CBCEFB}87.69 \\
     & T200k & 76.88 & 82.32 & 91.43 & 94.91 & 94.86 & 95.56 & \cellcolor[HTML]{CBCEFB}89.33 \\
\hline
\multirow{3}{*}{\QwenSeven{}} 
     & T50k & 71.11 & 79.83 & 87.12 & 90.73 & 91.82 & 90.58 & \cellcolor[HTML]{CBCEFB}85.20 \\
     & T100k & 75.15 & 81.33 & 90.20 & 93.54 & 94.05 & 93.98 & \cellcolor[HTML]{CBCEFB}88.04 \\
     & T200k & 76.83 & 81.81 & 92.03 & 95.19 & 95.15 & 95.38 & \cellcolor[HTML]{CBCEFB}89.40 \\
\hline
\multirow{3}{*}{\PaliGemma{}}  
     & T50k & 79.93 & 85.50 & 91.23 & 93.78 & 92.84 & 93.45 & \cellcolor[HTML]{CBCEFB}89.45 \\
     & T100k & 81.52 & 86.30 & 92.74 & 95.17 & 93.94 & 95.04 & \cellcolor[HTML]{CBCEFB}90.79 \\
     & T200k & 82.95 & 87.31 & 94.15 & 96.24 & 95.05 & 96.47 & \cellcolor[HTML]{CBCEFB}92.03 \\
\hline
\multirow{3}{*}{\GemmaFour{}} 
     & T50k & 80.49 & 85.95 & 92.76 & 95.33 & 95.79 & 96.44 & \cellcolor[HTML]{CBCEFB}91.12 \\
     & T100k & 82.39 & 86.92 & 94.67 & 96.71 & 96.71 & 97.30 & \cellcolor[HTML]{CBCEFB}92.45 \\
     & T200k & 83.63 & 87.63 & 95.93 & 97.42 & 97.23 & 98.02 & \cellcolor[HTML]{CBCEFB}93.31 \\
\hline \hline
\end{tabular}
\end{table}

\begin{table}[!htpb]\centering\footnotesize
\renewcommand{\arraystretch}{1.5}
\caption{Task~2 per-field Macro \(F_1\) (\%) on \test{}. Per-field values are unweighted over their allowed categories; Field mean is the unweighted average over the six fields.}
\label{tab:appendix_table_task2_2}
\begin{tabular}{cc rr rrrr c}
\hline \hline
\multirow{2}{*}{Model} & \multirow{2}{*}{Training set} & \multicolumn{2}{c}{Truth-label-derived} & \multicolumn{4}{c}{Rendered-image-derived} & \multirow{2}{*}{Field mean} \\
\cmidrule(lr){3-4} \cmidrule(lr){5-8}
 & & Family & Decay & Morph. & Comp. & Displ. & Charge & \\
\hline
\multirow{3}{*}{\QwenThree{}} 
     & T50k & 74.79 & 84.57 & 87.21 & 90.18 & 91.56 & 90.21 & \cellcolor[HTML]{CBCEFB}86.42 \\
     & T100k & 78.25 & 85.69 & 90.28 & 93.48 & 94.02 & 93.81 & \cellcolor[HTML]{CBCEFB}89.26 \\
     & T200k & 80.07 & 86.40 & 92.33 & 95.20 & 94.81 & 95.59 & \cellcolor[HTML]{CBCEFB}90.73 \\
\hline
\multirow{3}{*}{\QwenSeven{}}
     & T50k & 75.05 & 84.37 & 88.48 & 91.27 & 91.70 & 90.65 & \cellcolor[HTML]{CBCEFB}86.92 \\
     & T100k & 78.57 & 85.59 & 91.17 & 93.91 & 93.98 & 94.02 & \cellcolor[HTML]{CBCEFB}89.54 \\
     & T200k & 80.06 & 86.07 & 92.88 & 95.46 & 95.11 & 95.41 & \cellcolor[HTML]{CBCEFB}90.83 \\
\hline
\multirow{3}{*}{\PaliGemma{}} 
     & T50k & 81.16 & 87.83 & 91.56 & 93.81 & 92.84 & 93.44 & \cellcolor[HTML]{CBCEFB}90.10 \\
     & T100k & 82.65 & 88.43 & 93.02 & 95.19 & 93.94 & 95.04 & \cellcolor[HTML]{CBCEFB}91.38 \\
     & T200k & 83.94 & 89.21 & 94.40 & 96.24 & 95.05 & 96.47 & \cellcolor[HTML]{CBCEFB}92.55 \\
\hline
\multirow{3}{*}{\GemmaFour{}} 
     & T50k & 81.61 & 88.20 & 93.01 & 95.33 & 95.78 & 96.43 & \cellcolor[HTML]{CBCEFB}91.73 \\
     & T100k & 83.53 & 89.13 & 94.83 & 96.71 & 96.70 & 97.30 & \cellcolor[HTML]{CBCEFB}93.03 \\
     & T200k & 84.63 & 89.50 & 96.06 & 97.41 & 97.23 & 98.02 & \cellcolor[HTML]{CBCEFB}93.81 \\
\hline \hline
\end{tabular}
\end{table}

\section{Additional tables for Section~\ref{sec:scalinglaw_balanced} (Task~3)}\label{app:table_task3}
Tables~\ref{tab:appendix_table_task3_1} and \ref{tab:appendix_table_task3_2} provide the numerical Task~3 results underlying Figs.~\ref{fig:MT_task3_1} and \ref{fig:MT_task3_2}.

\begin{table}[!htpb]\centering\footnotesize
\renewcommand{\arraystretch}{1.0}
\caption{Task~3 binary consistency and replaced-panel localization results (\%) on \test{}. Binary Macro recall is the unweighted mean of the consistent- and inconsistent-example recalls; localization Macro recall is the unweighted mean over P1--P4 on the 12,000 inconsistent examples.}
\label{tab:appendix_table_task3_1}
\begin{tabular}{cc cccc}
\hline \hline
Model & Training set & Binary Macro recall & Consistent recall & Inconsistent recall & Localization Macro recall \\
\hline
\multirow{3}{*}{\QwenThree{}}
     & T50k & 94.60 & 97.42 & 91.78 & 89.10 \\
     & T100k & 98.15 & 99.04 & 97.26 & 96.52 \\
     & T200k & 99.26 & 99.62 & 98.89 & 98.60 \\
\hline
\multirow{3}{*}{\QwenSeven{}} 
     & T50k & 96.35 & 98.11 & 94.60 & 88.35 \\
     & T100k & 98.78 & 99.36 & 98.19 & 95.09 \\
     & T200k & 99.44 & 99.65 & 99.22 & 97.51 \\
\hline
\multirow{3}{*}{\PaliGemma{}} 
     & T50k & 98.10 & 99.03 & 97.16 & 96.56 \\
     & T100k & 99.15 & 99.52 & 98.78 & 98.45 \\
     & T200k & 99.61 & 99.84 & 99.38 & 99.23 \\
\hline
\multirow{3}{*}{\GemmaFour{}} 
     & T50k & 99.39 & 99.72 & 99.06 & 98.37 \\
     & T100k & 99.66 & 99.76 & 99.56 & 99.33 \\
     & T200k & 99.88 & 99.91 & 99.86 & 99.75 \\
\hline \hline
\end{tabular}
\end{table}

\begin{table}[!htpb]\centering\footnotesize
\renewcommand{\arraystretch}{1.0}
\caption{Task~3 replaced-panel localization recall (\%) on the 12,000 inconsistent examples in \test{}. Difficulty columns average equally over P1--P4, panel columns average equally over the three donor difficulty strata, and the final column gives localization Macro recall over P1--P4.}
\label{tab:appendix_table_task3_2}
\begin{tabular}{cc rrrrrrr c}
\hline \hline
\multirow{2}{*}{Model} & \multirow{2}{*}{Training set} & \multicolumn{3}{c}{Donor difficulty} & \multicolumn{4}{c}{Replaced panel} & \multirow{2}{*}{Localization Macro recall} \\
\cmidrule(lr){3-5}\cmidrule(lr){6-9}
 & & Easy & Medium & Hard & P1 & P2 & P3 & P4 & \\
\hline
\multirow{3}{*}{\QwenThree{}} 
     & T50k & 98.85 & 87.17 & 81.28 & 94.87 & 94.57 & 82.90 & 84.07 & \cellcolor[HTML]{CBCEFB}89.10 \\
     & T100k & 99.67 & 96.35 & 93.53 & 98.53 & 98.17 & 94.03 & 95.33 & \cellcolor[HTML]{CBCEFB}96.52 \\
     & T200k & 99.88 & 98.85 & 97.08 & 99.30 & 99.57 & 97.13 & 98.40 & \cellcolor[HTML]{CBCEFB}98.60 \\
\hline
\multirow{3}{*}{\QwenSeven{}}
     & T50k & 98.72 & 86.05 & 80.27 & 93.17 & 93.20 & 85.53 & 81.50 & \cellcolor[HTML]{CBCEFB}88.35 \\
     & T100k & 99.80 & 94.65 & 90.82 & 96.73 & 96.90 & 93.67 & 93.07 & \cellcolor[HTML]{CBCEFB}95.09 \\
     & T200k & 99.80 & 97.33 & 95.40 & 97.93 & 98.43 & 96.53 & 97.13 & \cellcolor[HTML]{CBCEFB}97.51 \\
\hline
\multirow{3}{*}{\PaliGemma{}}
     & T50k & 99.62 & 96.50 & 93.55 & 98.53 & 98.63 & 95.40 & 93.67 & \cellcolor[HTML]{CBCEFB}96.56 \\
     & T100k & 99.90 & 98.62 & 96.83 & 99.20 & 99.30 & 97.90 & 97.40 & \cellcolor[HTML]{CBCEFB}98.45 \\
     & T200k & 99.92 & 99.38 & 98.40 & 99.47 & 99.70 & 98.70 & 99.07 & \cellcolor[HTML]{CBCEFB}99.23 \\
\hline
\multirow{3}{*}{\GemmaFour{}}
     & T50k & 99.88 & 98.30 & 96.92 & 99.13 & 98.70 & 96.93 & 98.70 & \cellcolor[HTML]{CBCEFB}98.37 \\
     & T100k & 99.88 & 99.55 & 98.55 & 99.70 & 99.37 & 98.73 & 99.50 & \cellcolor[HTML]{CBCEFB}99.33 \\
     & T200k & 99.95 & 99.90 & 99.40 & 99.87 & 99.83 & 99.47 & 99.83 & \cellcolor[HTML]{CBCEFB}99.75 \\
\hline \hline
\end{tabular}
\end{table}

\end{appendices}

\end{document}